\documentstyle[12pt,epsfig,rotating] {article}
\begin{document}
\newcommand{\Z}{\mbox{$\mathrm{Z}$}}
\newcommand{\Zo}{\mbox{$\mathrm{Z^0}$}}
\newcommand{\W}{\mbox{$\mathrm{W}$}}
\newcommand{\bZo}{{\bf \mbox{$\mathrm{Z}$}}}
\newcommand{\Zg}{\mbox{$\mathrm{Z}^{0}\gamma$}}
\newcommand{\ZZ}{\mbox{$\mathrm{Z}^{0}\mathrm{Z}^{0}$}}
\newcommand{\WW}{\mbox{$\mathrm{W}\mathrm{W}$}}
\newcommand{\Zs}{\mbox{$\mathrm{Z}^{*}$}}
\newcommand{\h}{\mbox{$\mathrm{h}^{0}$}}
\newcommand{\Ho}{\mbox{$\mathrm{H}$}}
\newcommand{\ho}{\mbox{$\mathrm{h}$}}
\newcommand{\Hp}{\mbox{$\mathrm{H}^{+}$}}
\newcommand{\Hm}{\mbox{$\mathrm{H}^{-}$}}
\newcommand{\Hsm}{\mbox{$\mathrm{H}^{0}_{SM}$}}
\newcommand{\A}{\mbox{$\mathrm{A}^{0}$}}
\newcommand{\Hpm}{\mbox{$\mathrm{H}^{\pm}$}}
\newcommand{\X}{\mbox{${\tilde{\chi}^0}$}}
\newcommand{\ko}{\mbox{${\tilde{\chi}^0}$}}
\newcommand{\ee}{\mbox{$\mathrm{e}^{+}\mathrm{e}^{-}$}}
\newcommand{\bee}{\mbox{$\boldmath {\mathrm{e}^{+}\mathrm{e}^{-}} $}}
\newcommand{\mm}{\mbox{$\mu^{+}\mu^{-}$}}
\newcommand{\nn}{\mbox{$\nu \bar{\nu}$}}
\newcommand{\qq}{\mbox{$\mathrm{q} \bar{\mathrm{q}}$}}
\newcommand{\pb}{\mbox{$\mathrm{pb}^{-1}$}}
\newcommand{\ra}{\mbox{$\rightarrow$}}
\newcommand{\br}{\mbox{$\boldmath {\rightarrow}$}}
\newcommand{\tautau}{\mbox{$\tau^{+}\tau^{-}$}}
\newcommand{\ga}{\mbox{$\gamma$}}
\newcommand{\gamgam}{\mbox{$\gamma\gamma$}}
\newcommand{\tp}{\mbox{$\tau^+$}}
\newcommand{\tm}{\mbox{$\tau^-$}}
\newcommand{\tpm}{\mbox{$\tau^{\pm}$}}
\newcommand{\uu}{\mbox{$\mathrm{u} \bar{\mathrm{u}}$}}
\newcommand{\dd}{\mbox{$\mathrm{d} \bar{\mathrm{d}}$}}
\newcommand{\bb}{\mbox{$\mathrm{b} \bar{\mathrm{b}}$}}
\newcommand{\cc}{\mbox{$\mathrm{c} \bar{\mathrm{c}}$}}
\newcommand{\mumu}{\mbox{$\mu^+\mu^-$}}
\newcommand{\csbar}{\mbox{$\mathrm{c} \bar{\mathrm{s}}$}}
\newcommand{\cbars}{\mbox{$\bar{\mathrm{c}}\mathrm{s}$}}
\newcommand{\nunu}{\mbox{$\nu \bar{\nu}$}}
\newcommand{\nubar}{\mbox{$\bar{\nu}$}}
\newcommand{\mQ}{\mbox{$m_{\mathrm{Q}}$}}
\newcommand{\mZ}{\mbox{$m_{\mathrm{Z}}$}}
\newcommand{\mH}{\mbox{$m_{\mathrm{H}}$}}
\newcommand{\mHp}{\mbox{$m_{\mathrm{H}^+}$}}
\newcommand{\mh}{\mbox{$m_{\mathrm{h}}$}}
\newcommand{\mA}{\mbox{$m_{\mathrm{A}}$}}
\newcommand{\mHpm}{\mbox{$m_{\mathrm{H}^{\pm}}$}}
\newcommand{\mHsm}{\mbox{$m_{\mathrm{H}^0_{SM}}$}}
\newcommand{\mW}{\mbox{$m_{\mathrm{W}^{\pm}}$}}
\newcommand{\mt}{\mbox{$m_{\mathrm{t}}$}}
\newcommand{\mb}{\mbox{$m_{\mathrm{b}}$}}
\newcommand{\lpm}{\mbox{$\ell ^+ \ell^-$}}
\newcommand{\G}{\mbox{$\mathrm{GeV}$}}
\newcommand{\Gc}{\mbox{${\rm GeV}/c$}}
\newcommand{\Gcs}{\mbox{${\rm GeV}/c^2$}}
\newcommand{\Mcs}{\mbox{${\rm MeV}/c^2$}}
\newcommand{\sba}{\mbox{$\sin ^2 (\beta -\alpha)$}}
\newcommand{\cba}{\mbox{$\cos ^2 (\beta -\alpha)$}}
\newcommand{\tanb}{\mbox{$\tan \beta$}}
\newcommand{\sqrts}{\mbox{$\sqrt {s}$}}
\newcommand{\sqrtsp}{\mbox{$\sqrt {s'}$}}
\newcommand{\msusy}{\mbox{$M_{\rm SUSY}$}}
\newcommand{\mg}{\mbox{$m_{\tilde{\rm g}}$}}
\begin{titlepage}
\begin{center}
\vspace{-0.5cm}
{\Large EUROPEAN ORGANIZATION FOR NUCLEAR RESEARCH}
\end{center}
\bigskip
\begin{flushright}
ALEPH 2001-066 CONF 2001-046\\
DELPHI 2001-113 CONF 536\\
L3 Note 2699 \\
OPAL Physics Note PN479 \\
LHWG Note/2001-03\\
CERN-EP/2001-xxx\\
 {\today} 
\end{flushright}
\bigskip
\begin{center}{\Large \bf Search for the Standard Model Higgs Boson at LEP}
\end{center}
\begin{center}
      {\Large  ALEPH, DELPHI, L3 and OPAL Collaborations}\\
      \bigskip
      {\large The LEP working group for Higgs boson searches}
\end{center}
\bigskip
\begin{center}{\Large  Abstract}\end{center}
The four LEP collaborations, ALEPH, DELPHI, L3 and OPAL, have collected 
2465 pb$^{-1}$ of \ee\ collision data at energies between 189 and 209 GeV, of which 542~\pb\ 
were collected above 206~GeV.
Searches for the Standard Model Higgs boson have been performed by each of the LEP collaborations. 
Their data have been combined and examined for their consistency with the Standard Model background and various 
Standard Model Higgs boson mass hypotheses. 
A lower bound of 114.1~GeV has been obtained at the 95\% confidence level for the mass of the Higgs boson.
The likelihood analysis shows a preference for a Higgs boson with a mass of 115.6~GeV.
At this mass, the probability for the background
to generate the observed effect is 3.4\%.
\bigskip
\bigskip
\begin{center}
{\Large THE RESULTS QUOTED IN THIS PAPER ARE NOT FINAL}
\end{center}
\bigskip
\bigskip
\begin{center}
Contributed paper for EPS'01 in Budapest and LP'01 in Rome 
\end{center}
%
%
\end{titlepage}
\section{Introduction}
The Higgs mechanism~\cite{higgs} plays a central role in the unification of the electromagnetic and weak interactions
by providing mass to the intermediate vector bosons, \W\ and \Z, without violating local gauge invariance.
Within the Standard Model (SM)~\cite{sm}, the Higgs mechanism predicts a single neutral scalar particle, the
Higgs boson. Its mass is arbitrary; however, self-consistency of the model up to a scale $\Lambda$ imposes
an upper~\cite{upper-bound} and lower bound~\cite{lower-bound}. If $\Lambda$ is close to the Planck scale, the mass
of the SM Higgs boson is confined between about 130 and 190~GeV~\cite{bounds}. A mass less than 130~GeV would
indicate physics beyond the SM to set in below the Planck scale; for example, in the minimal supersymmetric extension 
of the SM the mass of the lightest neutral scalar \h\ is predicted to be less than 135~GeV~\cite{mssm}.
Even stronger bounds are obtained using arguments of naturalness and fine-tuning~\cite{fine-tuning}.

Indirect experimental constraints are derived from precision measurements of 
electroweak parameters which depend in their interpretation on the $\log$ of the Higgs boson mass via radiative corrections. 
If the SM is assumed, the currently preferred mass value is $m_{\rm H}= 88 ^{+53}_{-33}$~GeV, and the 
95\% confidence level upper bound on the mass is 196~GeV~\cite{blueband}.    

Direct searches carried out by the four LEP collaborations in data collected prior to the year 2000 did not reveal any
signal for the SM Higgs boson. When the LEP data were statistically added, the observed event
rate and their distributions have shown good agreement with the SM background processes~\cite{adlo-cernep,adlo-osaka, osaka}.

The situation changed during summer 2000 with the advent of new LEP data, at centre-of-mass energies exceeding 206~GeV.
At the session of the LEP Committee of September 5, 2000, ALEPH reported an excess of events suggesting the production 
of a SM Higgs boson with mass in the vicinity of 
115~GeV~\cite{aleph-septlepc} while DELPHI, L3 and OPAL did not support this observation.
The quoted  probabilities for the SM background to produce the observed event configuration
($1-CL_b$, as defined below in Section 2.5), are listed 
in the first line of Table~\ref{tab:history} where the LEP combined result is also quoted.
Due to this ambiguous situation, the LEP shutdown planned for the end of September was postponed by one month, 
and all effort was made to maximize the LEP energy.  
\begin{table}[htb]
\begin{center}
\begin{tabular}{|l|cccc|c|}
\hline
              & ALEPH                & DELPHI     & L3        & OPAL    & LEP    \\
\hline\hline
LEPC, Sept 5 $(^*)$ & $1.6\times 10^{-4}$  & 0.67       & 0.84      & 0.47    & $2.5\times 10^{-2}$ \\
LEPC, Nov 3 \cite{pik-lepc}  & $6.5\times 10^{-4}$  & 0.68       & 0.068     & 0.19    & $4.2\times 10^{-3}$ \\
Ref's~\cite{aleph,delphi,l3-new,opal} & $2.6\times 10^{-3}$  & 0.77       & 0.32      & 0.20    &    \\ 
\hline
\end{tabular}
\caption{\small Background probabilities ($1-CL_b$) at a Higgs boson test-mass of $m_H=115$~GeV, for the individual
experiments and for the LEP data combined. $(^*)$ The results presented at the Sept. 5 LEPC have been revised for the LEPC of
Nov. 3. The values listed are the revised ones~\cite{pik-lepc}.
\label{tab:history}}
\end{center}
\end{table}

A rapid analysis which included the bulk part of the new data
resulted in the probabilities listed in the second line of Table~\ref{tab:history}.
These results were presented at the LEP Committee meeting of November 3, 2000~\cite{pik-lepc}.
The ALEPH excess was slightly attenuated, as indicated by the increased background probability. 
On the other hand, L3 reported some
candidates suggesting the Higgs boson interpretation~\cite{l3}.

After a thorough review of the analysis procedures, the LEP collaborations have published their
results~\cite{aleph,delphi, l3-new,opal}, updating them to include all data. The L3 publication~\cite{l3-new} 
is final. 
The review addressed many potential systematic
errors, especially in the handling of a signal at the kinematic limit of the production process
\ee\ra~ZH. Also, the uncertainties from Monte
Carlo statistics were reduced and in some cases the search sensitivity has been improved. 
The published background probabilities at a test-mass of 115~GeV 
are reported in the last line of Table~\ref{tab:history}. The  ALEPH~\cite{aleph} and L3~\cite{l3-new}
excesses have decreased since the beginning of November.

In this paper we present combined results from LEP which are based on these recent publications. 
However, the inputs also include data collected before the year 2000.
The c.m. energies ($E_{cm}$) thus span the range from 189~GeV to 209~GeV.
The integrated luminosities by experiment and energy are given in Table~\ref{tab:lumi}. 
\begin{table}[htb]
\begin{center}
\begin{tabular}{|l|cccc|c|}
\hline
                          & ALEPH      & DELPHI     & L3        & OPAL   & LEP    \\
\hline\hline
$E_{cm} \ge 189$ GeV      & 629        & 610        & 627       & 599    & 2465   \\
$E_{cm} \ge 206$ GeV      & 130        & 142        & 139       & 130    & 542    \\
\hline
\end{tabular}
\caption{\small Integrated luminosities (pb$^{-1}$) of the data samples provided by the four experiments 
for the present combination, and of the total LEP sample. Subsets taken at energies larger than 206~GeV 
are listed separately. 
\label{tab:lumi}}
\end{center}
\end{table}

At LEP energies, the SM Higgs boson is expected to be produced mainly in association with a \Z\ boson through the 
Higgsstrahlung process,
\ee\ra\Ho\Z~\cite{bjorken}. Small additional contributions are expected from t-channel \W\ and \Z\ boson fusion 
processes, which produce a Higgs boson and either a pair of neutrinos or electrons in the final state~\cite{fusion}.
For masses in the vicinity of 115~GeV (the kinematic limit for Higgsstrahlung at $E_{cm}\approx 206$~GeV), the SM Higgs boson is expected to decay mainly into \bb\ quark pairs
(74\%) while decays to tau lepton pairs, WW$^*$, gluon pairs ($\approx$ 7\% each), and to \cc\ ($\approx$ 4\%)
are all less important. 
The final-state topologies are determined by these decays and by the decay of the associated \Z\ boson. 
The searches at LEP encompass the  four-jet final state (\Ho\ra\bb)\qq, the missing
energy final state (\Ho\ra\bb)\nn, the leptonic final state (\Ho\ra\bb)$\ell^+\ell^-$ where $\ell$ denotes an 
electron or a muon, and the tau
lepton final states (\Ho\ra\bb)$\tau^+\tau^-$ and (\Ho\ra$\tau^+\tau^-$)(\Z\ra\qq).
 
Preselection cuts are applied to reduce the main background from two-photon processes and from
radiative returns to the \Z\ boson, \ee\ra\Z$\gamma(\gamma)$. 
The remaining background, mainly from fermion pairs (possibly with photon or gluon radiation), \WW, and ZZ, is reduced by applying cuts
which make use of kinematic differences between the signal and the background processes and of the requirement 
of b-flavour, abundant in the decay of the Higgs boson. The detailed implementation of these selections 
and analysis procedures is different for
each experiment~\cite{aleph}-\cite{opal}. In some search channels
\footnote{In the following, the word channel designates any subset 
of the data where the Higgs boson search is carried out; these may correspond to different final state topologies, 
to subsets of data 
collected at different c.m. energies or to subsets provided by different experiments.},
the selection 
depends explicitly upon the hypothesized Higgs boson mass.
\section{Combination procedure and results}
\subsection{Input provided by the experiments}
The information provided by the LEP experiments as input to the combination  
is in most cases binned in two 
discriminating variables:  (i) the reconstructed
Higgs boson mass $m_H^{rec}$, and  (ii) a variable $\cal G$ which combines many features of the events and 
allows the analysis to distinguish on a statistical basis between events from the Higgs boson signal 
and events from background 
processes. This variable 
is typically the outcome of a likelihood analysis or the output of an artificial neural network. 
Variables which tag b-flavoured jets contribute in an essential way to 
the value of $\cal G$. 

In a given bin $i$ of the plane defined by $m_H^{rec}$ and $\cal G$, the experiments provide the number $N_i$ of selected 
data events, the expected background rate $b_i$, and the expected signal $s_i(m_H)$
for a set of hypothesized 
Higgs boson masses (test-mass $m_H$ hereafter). 
In those channels where the selection depends on
$m_H$, the values of $N_i$ and $b_i$ are also given for a set of $m_H$ values. For a given test-mass, 
a weight of $s/b$ can thus be assigned to each selected candidate, depending on $m_H$ and the bin
where it is reconstructed.
The estimation of $s_i$ and $b_i$  makes use of detailed Monte Carlo simulations which take into account all known 
experimental features such as the c.m. energy and integrated luminosity of the data samples, cross-sections and 
decay branching ratios for the signal and background processes,  
selection efficiencies, experimental resolutions with non-gaussian contributions and systematic errors with 
their correlations. 
Since the simulation is done at fixed sets of
$E_{cm}$ and $m_H$, interpolation procedures such as~\cite{interpol} are applied to obtain
the distributions which correspond to arbitrary energies and test-masses. In order to avoid problems which might arise
in some bins due to low Monte Carlo statistics, smoothing procedures such as~\cite{keys} are applied which use the
corresponding information in the neighbouring bins. 
\subsection{Hypothesis testing}
The observed data configuration in the [$m_H^{rec},~\cal G$] plane  is subjected to a likelihood test
of two hypothetical scenarios. In the background scenario 
it is assumed that the data receive contributions from the SM background processes only while in the signal+background
scenario the contribution from a Higgs boson of test-mass $m_H$ is assumed in addition. The expressions for the corresponding 
likelihoods, ${\cal L}_{b}$ and ${\cal L}_{s+b}$, are given e.g. in Appendix A of Ref.~\cite{adlo-cernep}. The ratio
\begin{equation}
         Q= {\cal L}_{s+b}/ {\cal L}_{b}
\end{equation}
serves as test-statistic allowing to rank any data configuration between the background and signal+background hypotheses. 
For convenience, the quantity
\begin{equation}
-2\ln Q = 2s_{tot} - 2\sum_{i} N_i\ln[1+s_i/b_i]
\end{equation}
is used since in the limit of high statistics it corresponds to the difference in $\chi^2$ between 
the two hypotheses. In the above expression, $s_{tot}=\sum_{i} s_i$ is the total expected signal rate.
This test-statistic has been adopted since it makes the most efficient use of the information available in the observed
event configuration of a search, similarly to the way the principle of maximum likelihood gives the most efficient 
estimators of parameters in a measurement. 

\begin{figure}[htb]
\begin{center}
\epsfig{figure=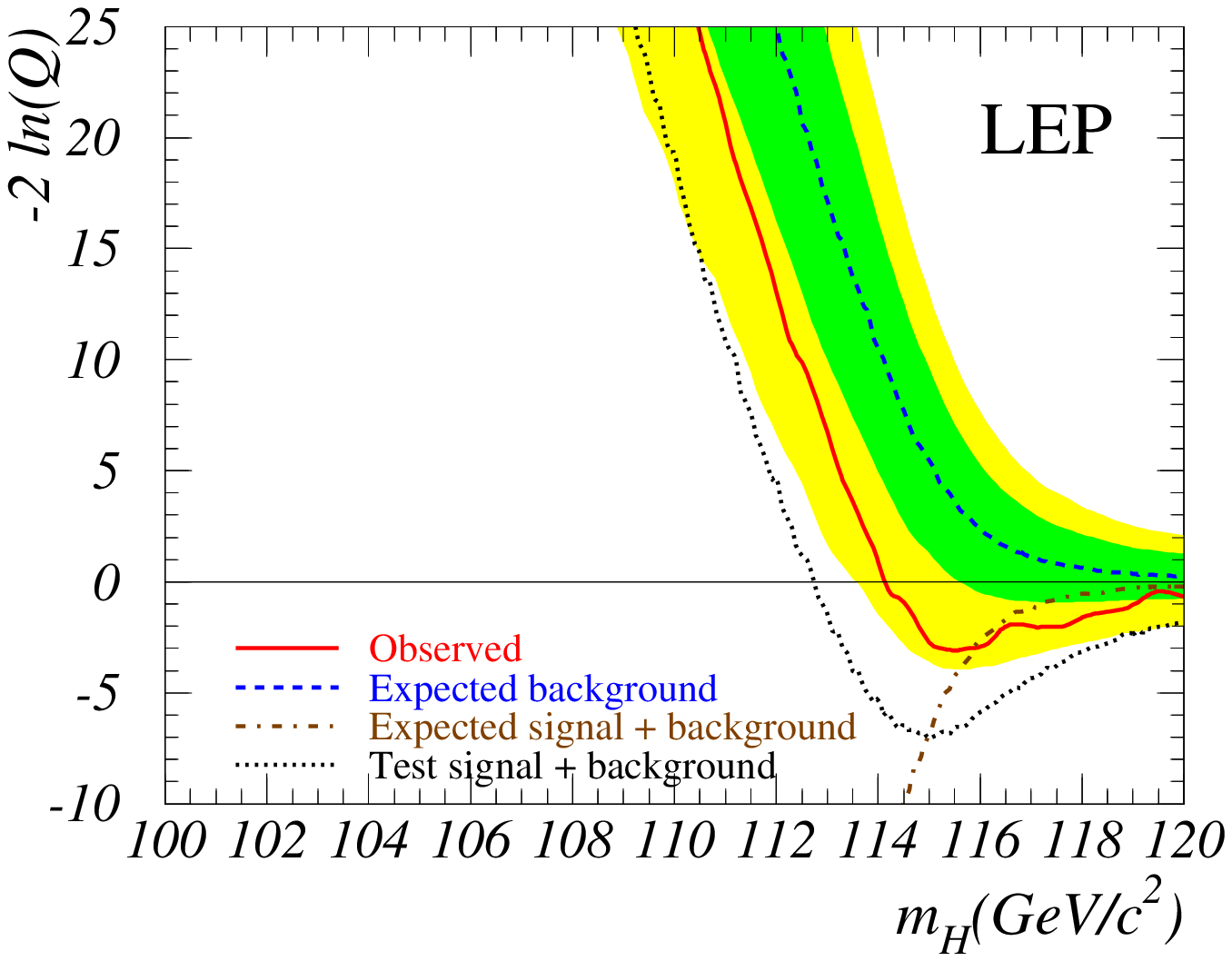,width=\textwidth}\\
\caption[]{\small  Observed and expected behaviour of the likelihood ratio $-2\ln Q$ as a function of the test-mass
$m_H$, obtained by combining the data of all four experiments. The solid line represents the observation; the dashed/dash-dotted lines show the median 
background/signal+background expectations. The dark/light shaded bands around the background expectation represent the 
$\pm 1$/$\pm 2$ standard deviation spread of the background expectation obtained from a large number of background experiments.
The dotted line is the result of a test where the signal from a 115 GeV Higgs boson has been
added to the background and propagated through the likelihood ratio calculation.
\label{fig:adlo-lnq}}
\end{center}
\end{figure}

Figure~\ref{fig:adlo-lnq} shows the test-statistic $-2\ln Q$ 
as a function of the test-mass for the present combination of LEP data.
The expected curves and their spreads
are obtained by replacing the observed data configuration by a large number of simulated event 
configurations. 
 
There is a minimum in the observed $-2\ln Q$ at $m_H=115.6$~GeV (maximum of the likelihood ratio $Q$)
indicating a deviation from the background 
hypothesis. The minimum coincides with the signal+background expectation for the same test-mass.
The value of $-2\ln Q$ at $m_H=115.6$~GeV is $-2.88$. 

Another feature in Figure~\ref{fig:adlo-lnq} is a persistent tail in the observation
towards lower test-masses where the observed curve stays away from the prediction for background. This is interpreted 
as being due to a large extent to the experimental resolution. A test has been performed  where the signal expected from a 
115~GeV Higgs boson was injected in the background simulation and propagated through the likelihood ratio calculation
at each $m_H$ value.
Although the resulting curve (dotted line) reproduces the main feature of the observed tail
\footnote{For a Higgs mass of 115.6~GeV, the outcome would follow closely the dotted curve, slightly 
displaced, so that its minimum coincides with the signal+background expectation (dash-dotted curve) at $m_H=115.6$~GeV.},
local excess events due
to statistical fluctuations can also contribute to the tail.
 
\begin{figure}[htb]
\begin{center}
\epsfig{figure=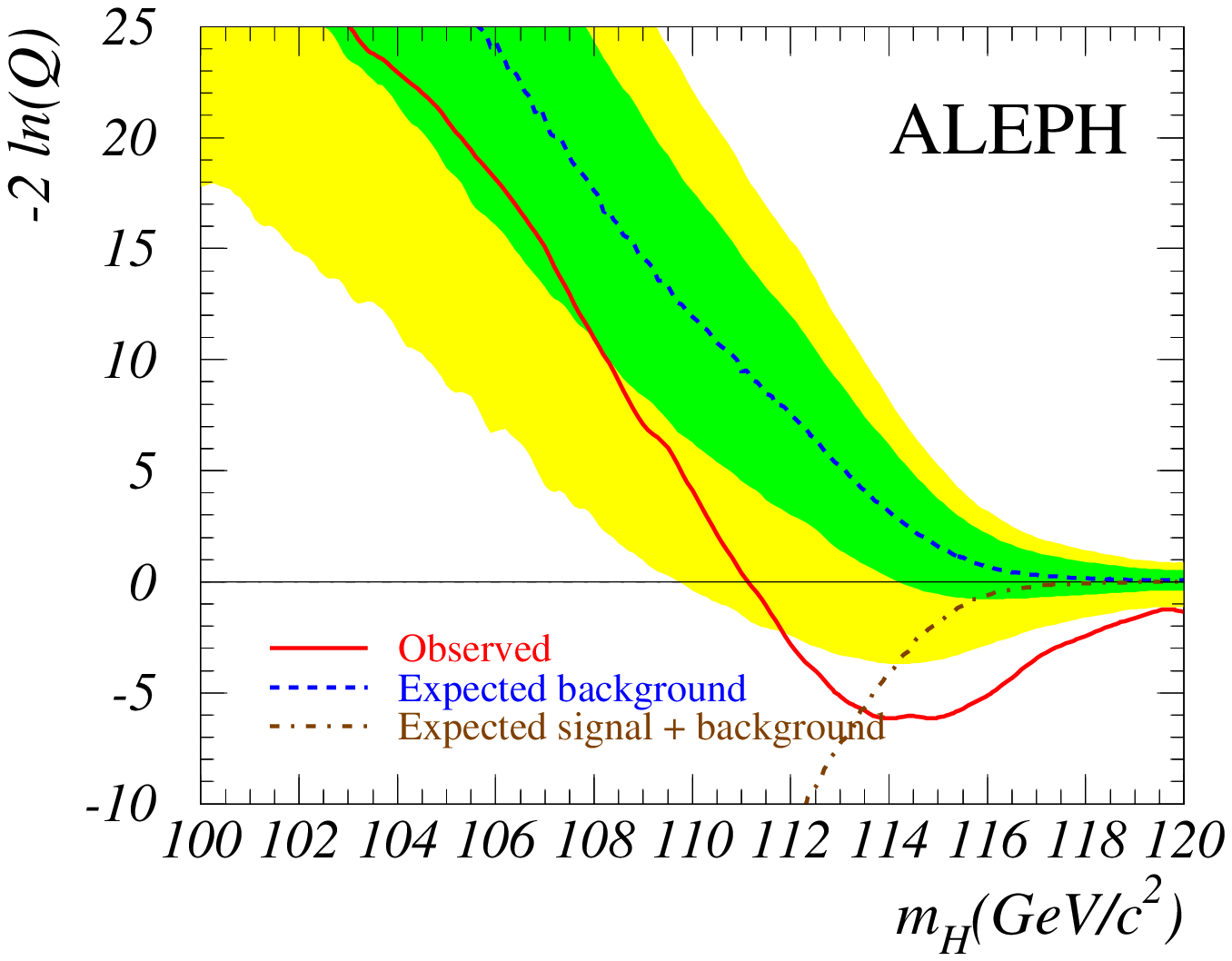,width=0.49\textwidth}
\epsfig{figure=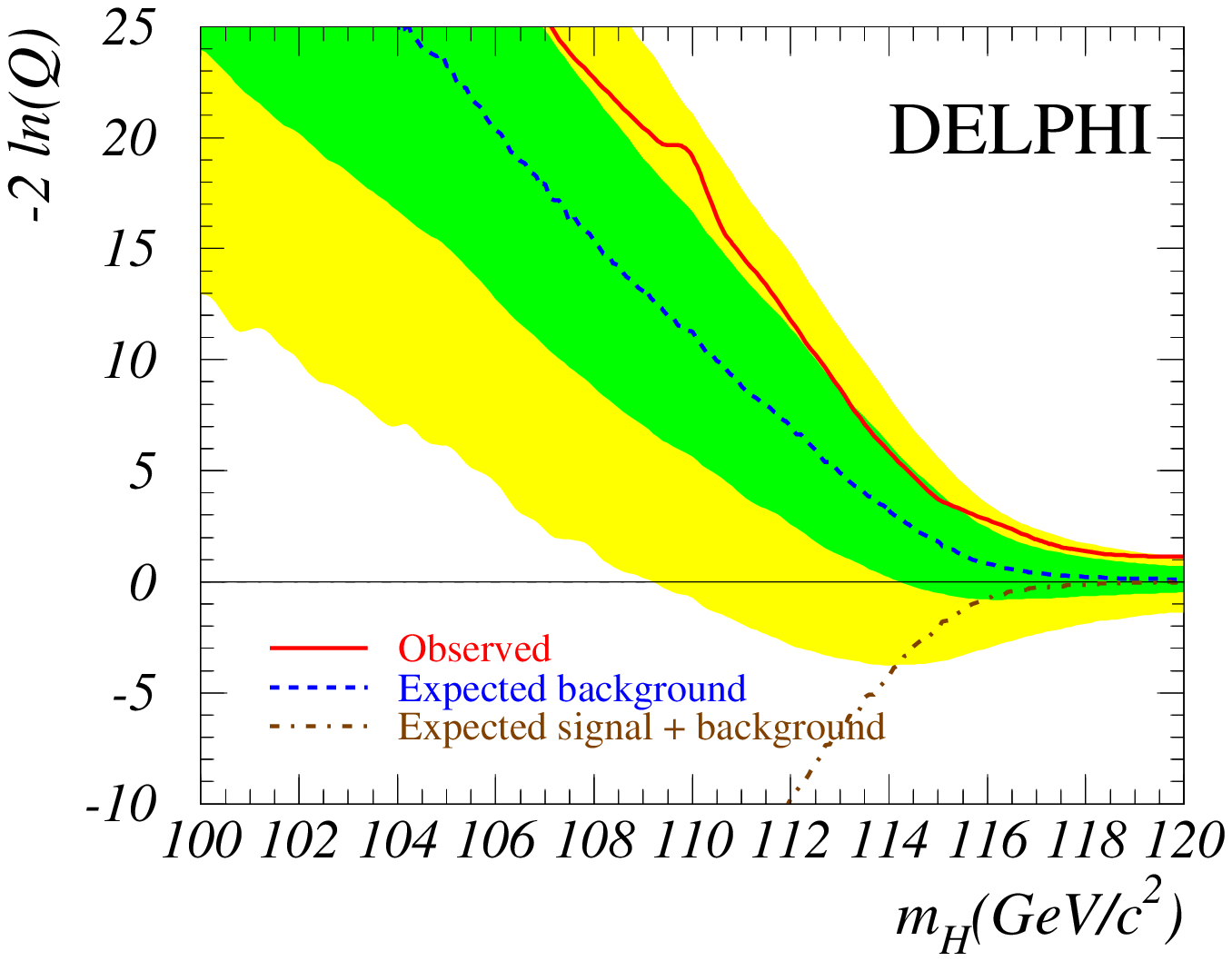,width=0.49\textwidth}\\
\epsfig{figure=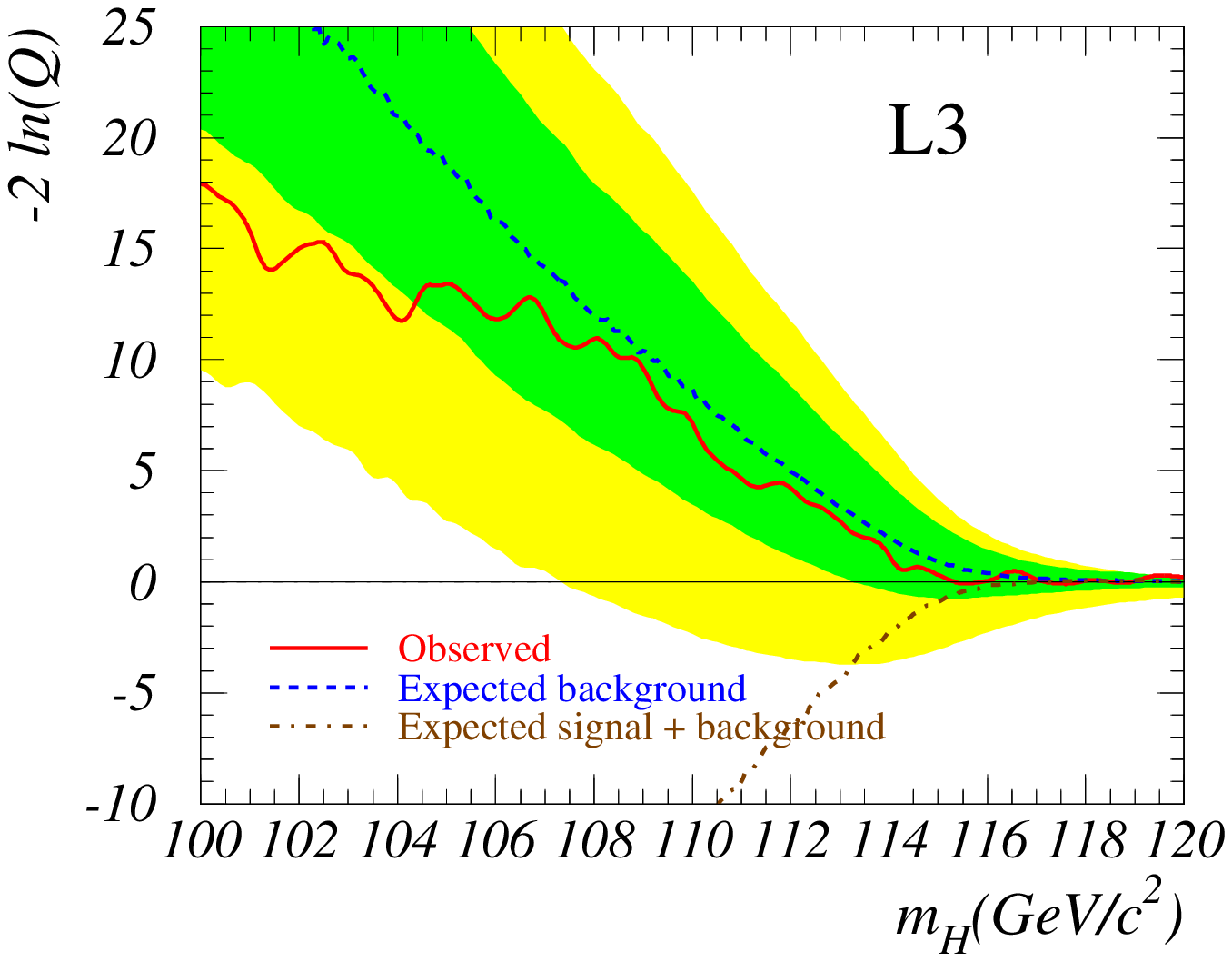,width=0.49\textwidth}
\epsfig{figure=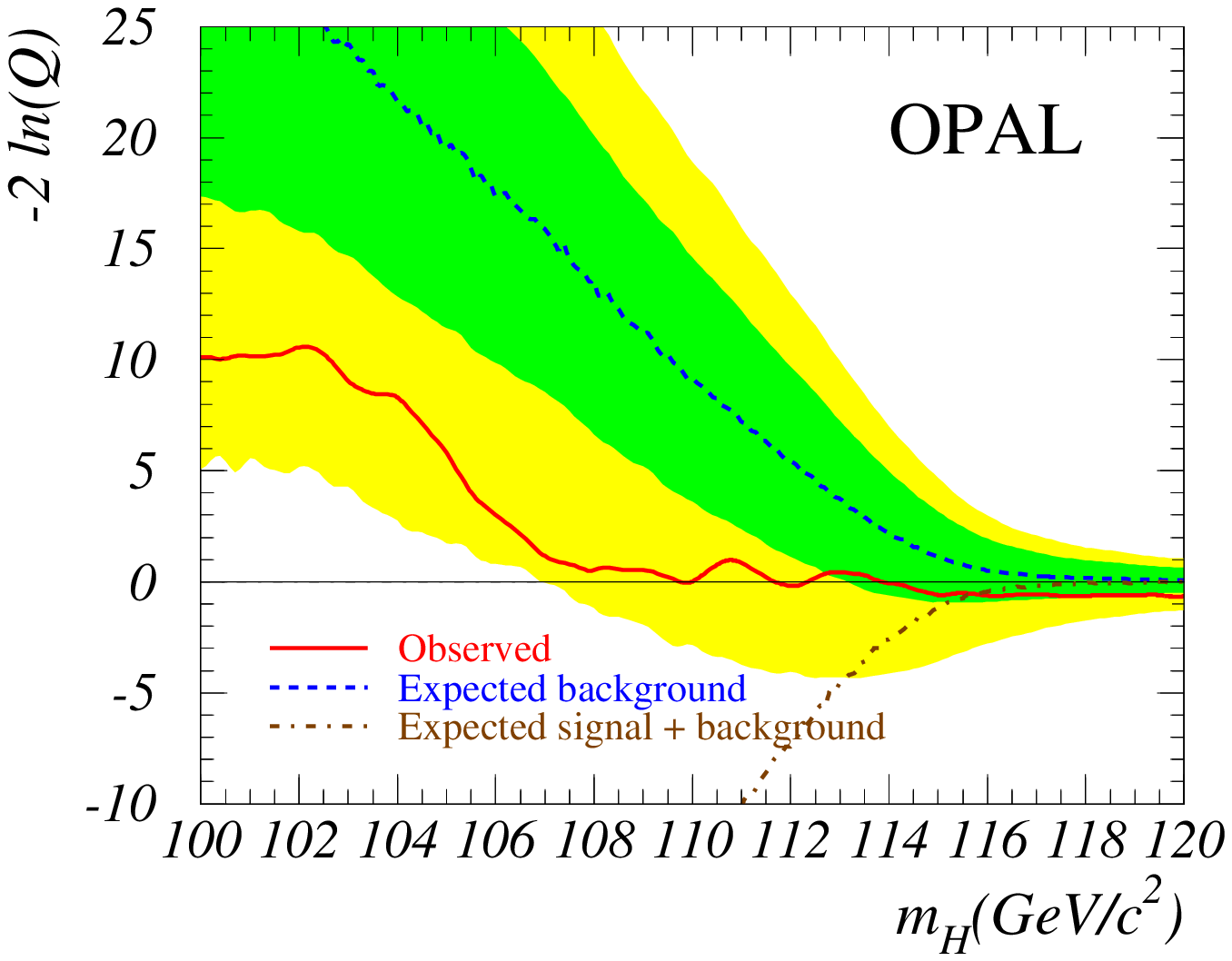,width=0.49\textwidth}\\
\caption[]{\small Observed and expected behaviour of the test statistic ($-2\ln Q$) as a function of the test-mass
$m_H$ obtained when the combination procedure is applied to the data sets from single experiments 
(see Figure~\ref{fig:adlo-lnq} for the notations).
\label{fig:a-d-l-o-lnq}}
\end{center}
\end{figure}

\begin{figure}[htb]
\begin{center}
\epsfig{figure=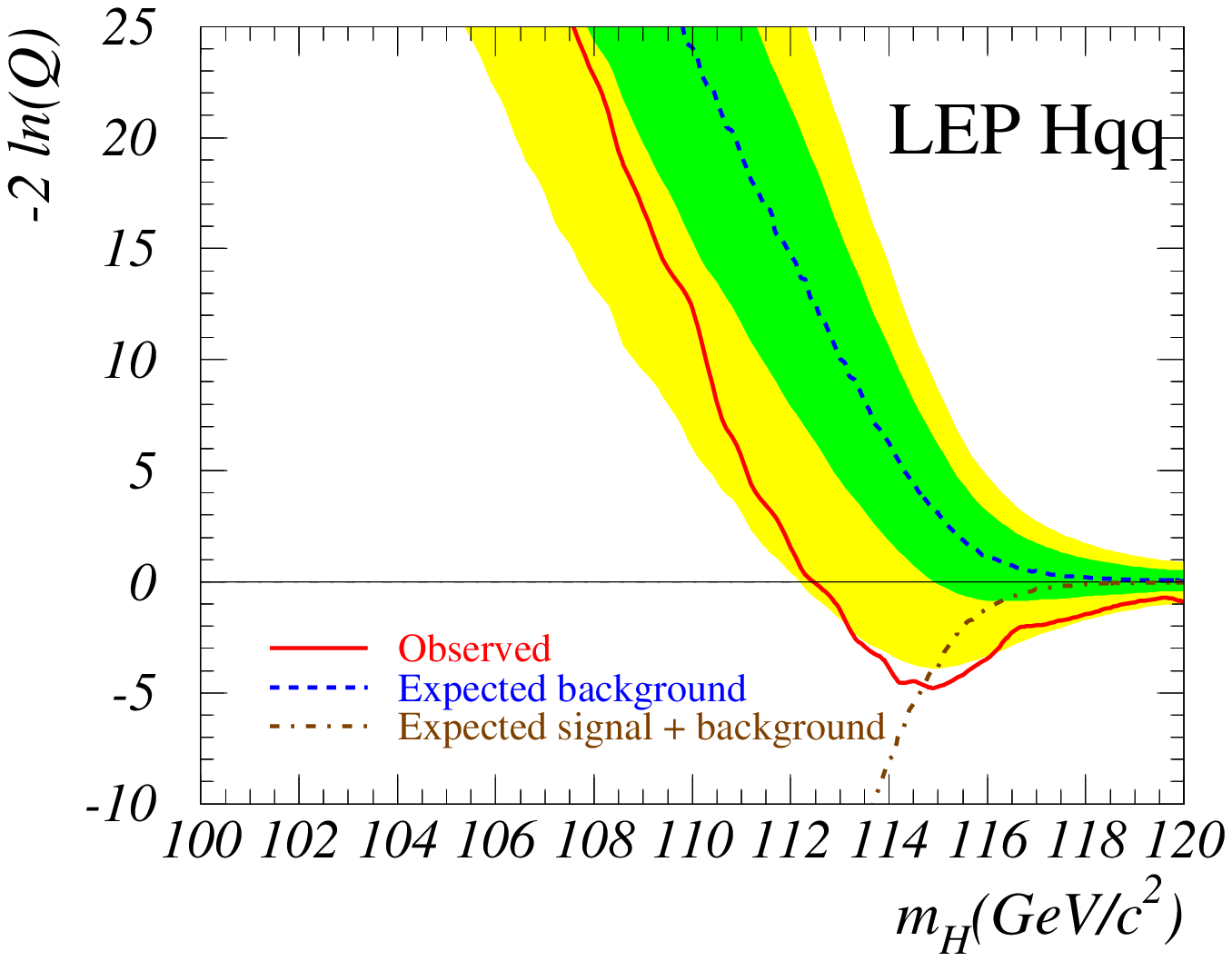,width=0.49\textwidth}
\epsfig{figure=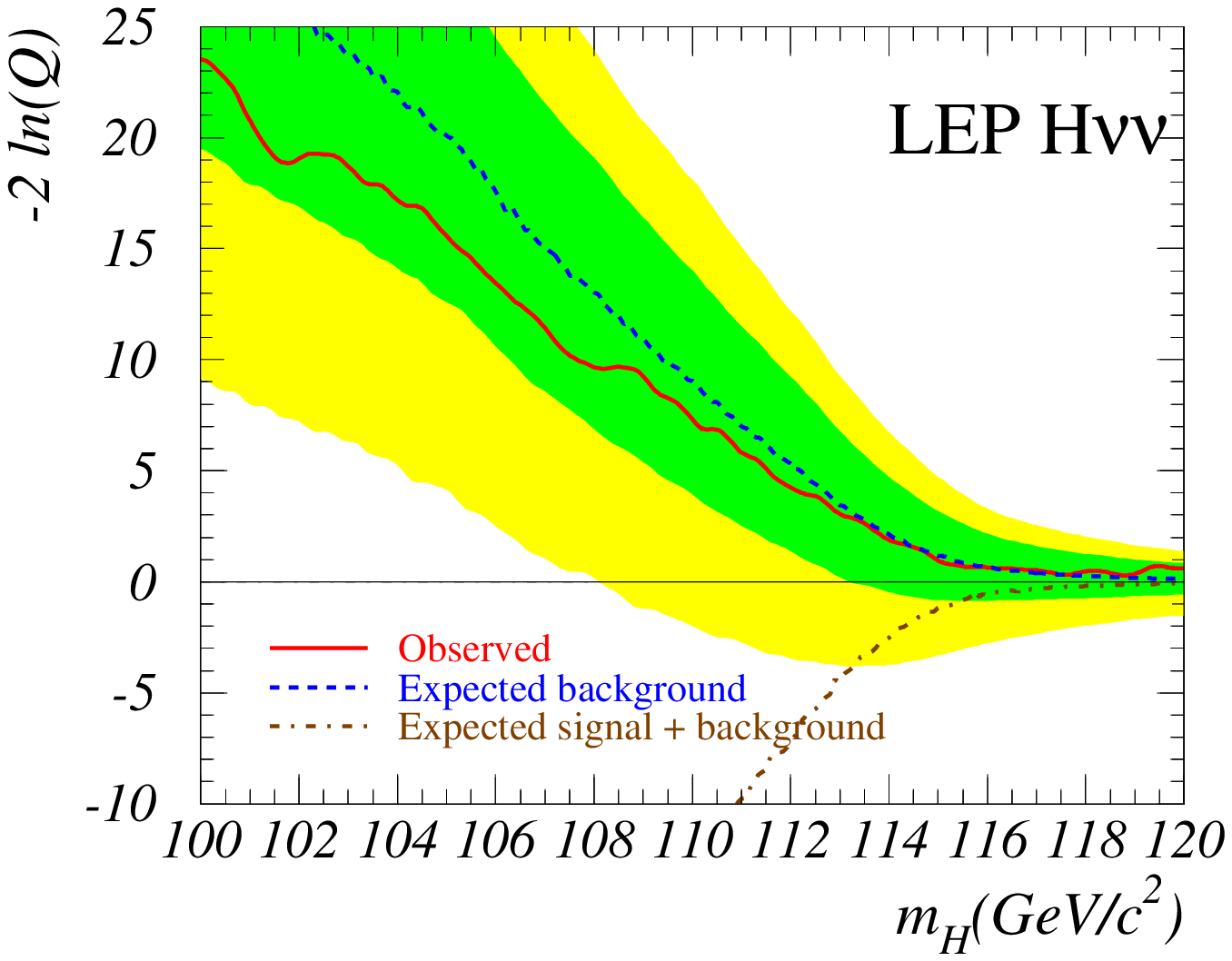,width=0.49\textwidth}\\
\epsfig{figure=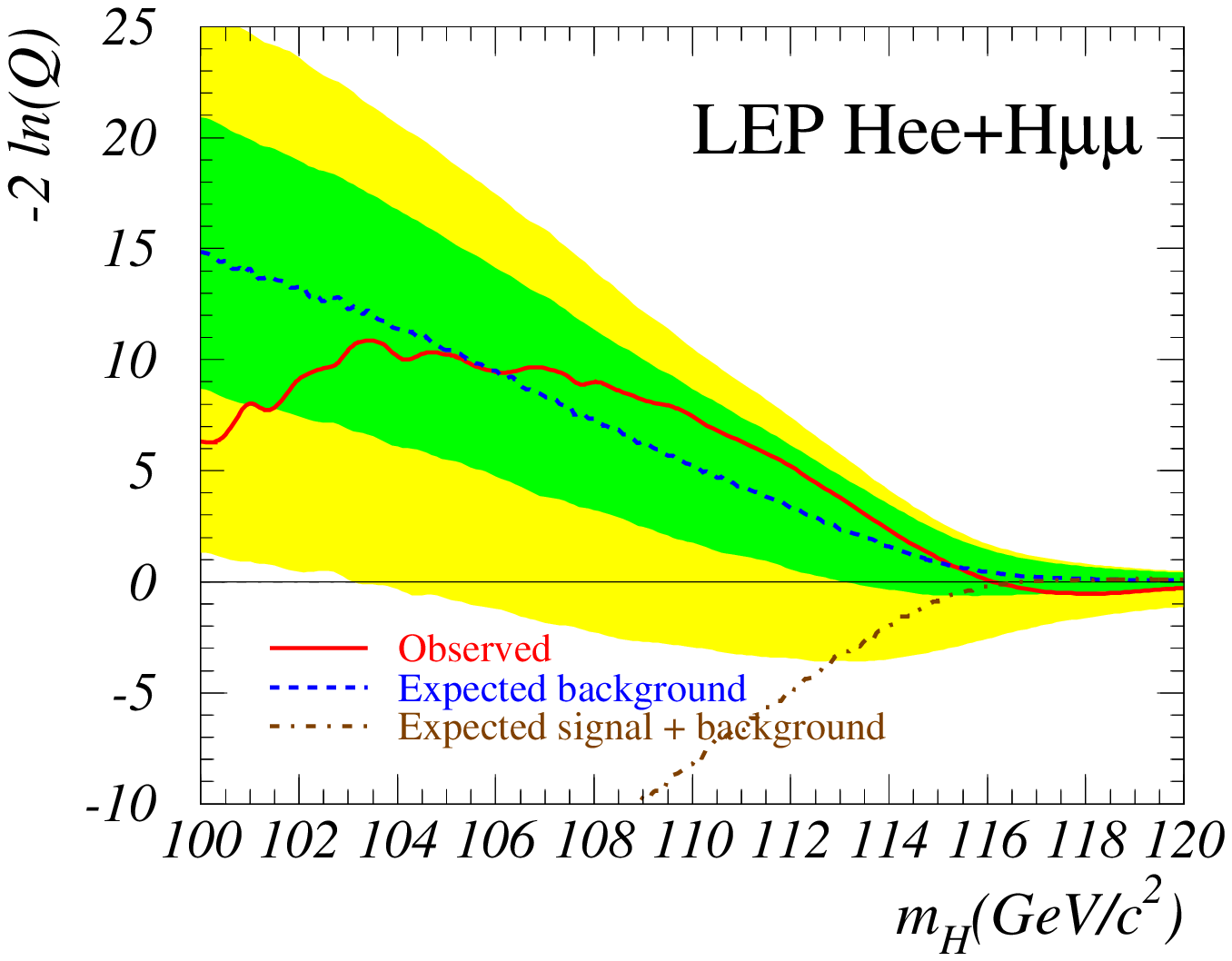,width=0.49\textwidth}
\epsfig{figure=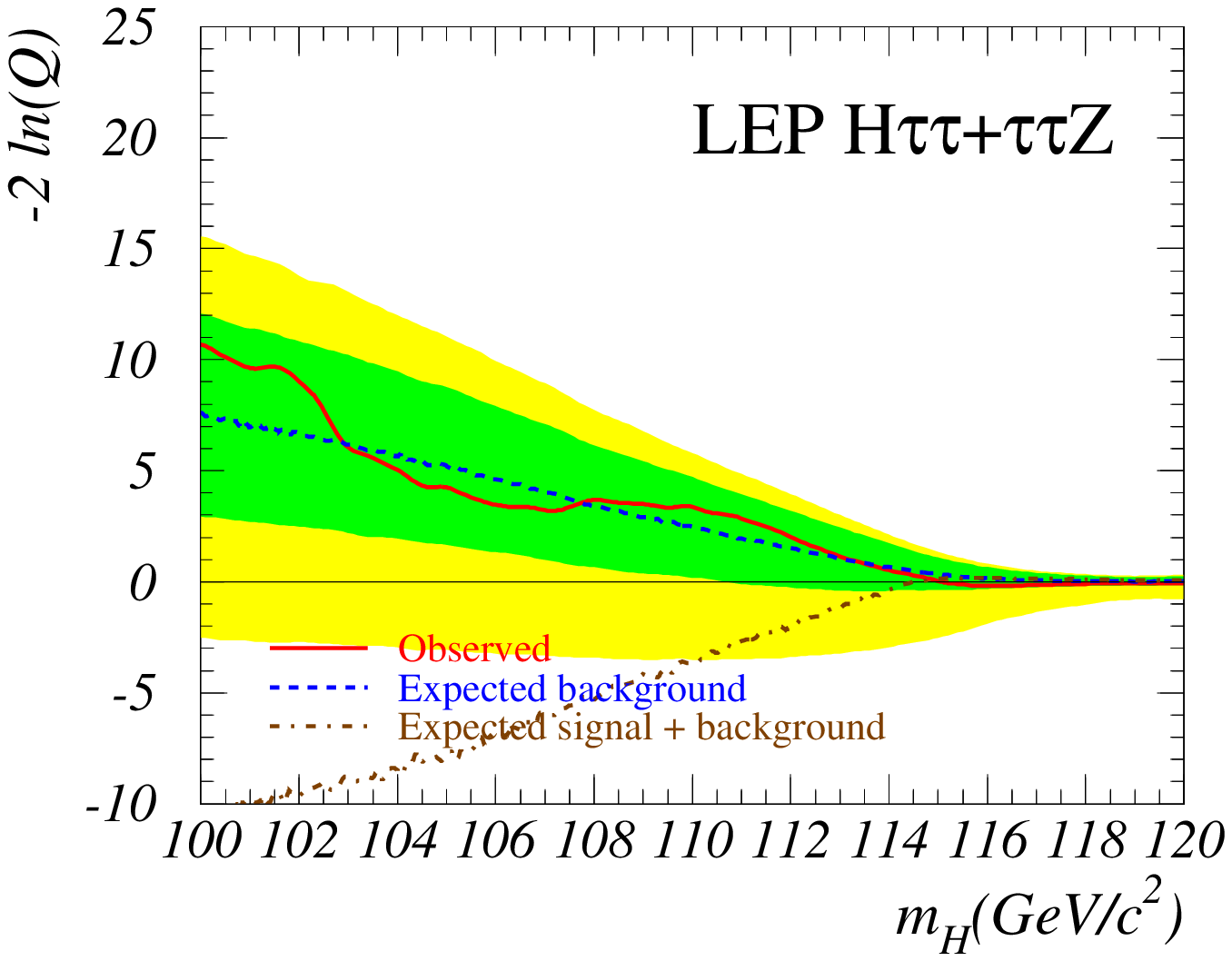,width=0.49\textwidth}\\
\caption[]{\small  
Observed and expected behaviour of the test statistic ($-2\ln Q$) as a function of the test-mass
$m_H$ obtained when the combination procedure is applied to the inputs corresponding to separated decay channels
(see Figure~\ref{fig:adlo-lnq} for the notations).
\label{fig:channels-lnq}}
\end{center}
\end{figure}

In Figures~\ref{fig:a-d-l-o-lnq} and~\ref{fig:channels-lnq}
the likelihood test is applied to subsets of the data, from individual experiments and final-state topologies.
In the vicinity of $m_H=115$~GeV, the signal-like behaviour mainly
originates from the ALEPH data and is concentrated in the four-jet final state.
One should note that none of the four experiments, taken separately, have the statistical power to distinguish 
between the background and the signal+background hypotheses at the level of two standard deviations 
for a test mass of 115~GeV (see the intersection of the signal+background curve with the lower edge of the light-shaded bands).
Among the final-state topologies, only the LEP combined four-jet channel is sufficiently
powerful to do so. 
\subsection{Contributions from single events}
\begin{figure}[htb]
\begin{center}
\epsfig{figure=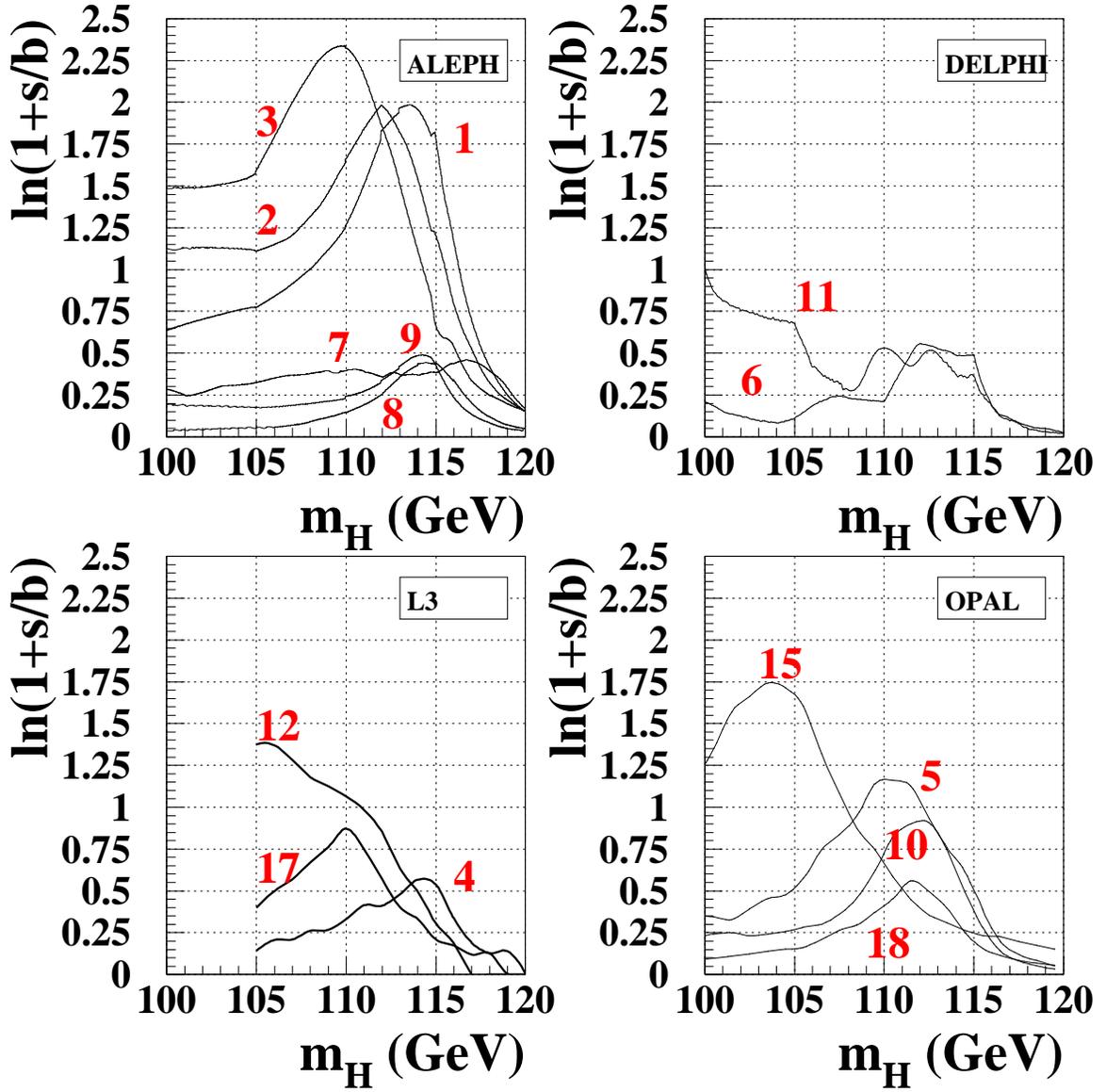,width=\textwidth}\\
\caption[]{\small 
Evolution of the event weight $\ln (1+s/b)$
with test-mass $m_H$, for the events with the largest contributions to $-2\ln Q$ at $m_H=115$~GeV. 
The labels correspond to the numbering in the first column of Table~\ref{tab:event-list}.
\label{fig:spaghetti}}
\end{center}
\end{figure}

The likelihood ratio $-2\ln Q$ is built up from individual event weights $\ln(1+s/b)$. 
The 20 candidates with the highest weights at $m_H=115$~GeV  are listed in Table~\ref{tab:event-list}. 
Some of these candidates are discussed in detail in Ref's~\cite{aleph}, \cite{l3}, \cite{l3-new}, \cite{opal}
and~\cite{l3-note}. 
\begin{table}[htb]
\begin{center}
\begin{tabular}{|c|lcc|cc|}
\hline
   & Expt&  $E_{cm}$ &  Decay channel &   $M_H^{rec}$ (GeV) &  $\ln (1+s/b)$ @115 GeV\\
\hline\hline
1&  Aleph  &  206.7 &  4-jet &   114.3   & 1.73\\
2&  Aleph  &  206.7 &  4-jet &   112.9   & 1.21\\
3&  Aleph  &  206.5 &  4-jet &   110.0   & 0.64\\
4&  L3     &  206.4 &  E-miss&   115.0   & 0.53\\
5&  Opal   &  206.6 &  4-jet &   110.7   & 0.53\\
6&  Delphi &  206.7 &  4-jet &   114.3   & 0.49\\
7&  Aleph  &  205.0 &  Lept  &   118.1   & 0.47\\
8&  Aleph  &  208.1 &  Tau   &   115.4   & 0.41\\
9&  Aleph  &  206.5 &  4-jet &   114.5   & 0.40\\
10& Opal   &  205.4 &  4-jet &   112.6   & 0.40\\
11& Delphi &  206.7 &  4-jet &    97.2   & 0.36\\
12& L3     &  206.4 &  4-jet &   108.3   & 0.31\\
13& Aleph  &  206.5 &  4-jet &   114.4   & 0.27\\
14& Aleph  &  207.6 &  4-jet &   103.0   & 0.26\\
15& Opal   &  205.4 &  E-miss&   104.0   & 0.25\\ 
16& Aleph  &  206.5 &  4-jet &   110.2   & 0.22\\
17& L3     &  206.4 &  E-miss&   110.1   & 0.21\\
18& Opal   &  206.4 &  E-miss&   112.1   & 0.20\\
19& Delphi &  206.7 &  4-jet &   110.1  & 0.20\\ 
20& L3     &  206.4 &  E-miss&   110.1   & 0.18\\
\hline 
\end{tabular}
\caption{\small Properties of the 20 candidates contributing with the highest weight $\ln (1+s/b)$ to  
$-2\ln Q$ at $m_H=115$~GeV. The experiment, c.m. energy, decay channel, the reconstructed mass and the weight 
at $m_H=115$~GeV are listed. This list is obtained requiring $s/b > 0.2$ or $\ln(1+s/b > 0.18$ at $m_H=115$~GeV.
The corresponding expected signal and background rates are 8.8 and 16.5 events, respectively.
\label{tab:event-list}}
\end{center}
\end{table}
For the events of each experiment with the highest weight at $\mH=115$~GeV,
the evolution of $\ln(1+s/b)$ with test-mass is shown in Figure~\ref{fig:spaghetti}. 
Due to the experimental resolution, candidate events with a
given reconstructed mass are seen to have sizeable weights for a range of test-masses, with the maximum weight being for 
test-masses close to the reconstructed mass.

\begin{figure}[htb]
\begin{center}
\epsfig{figure=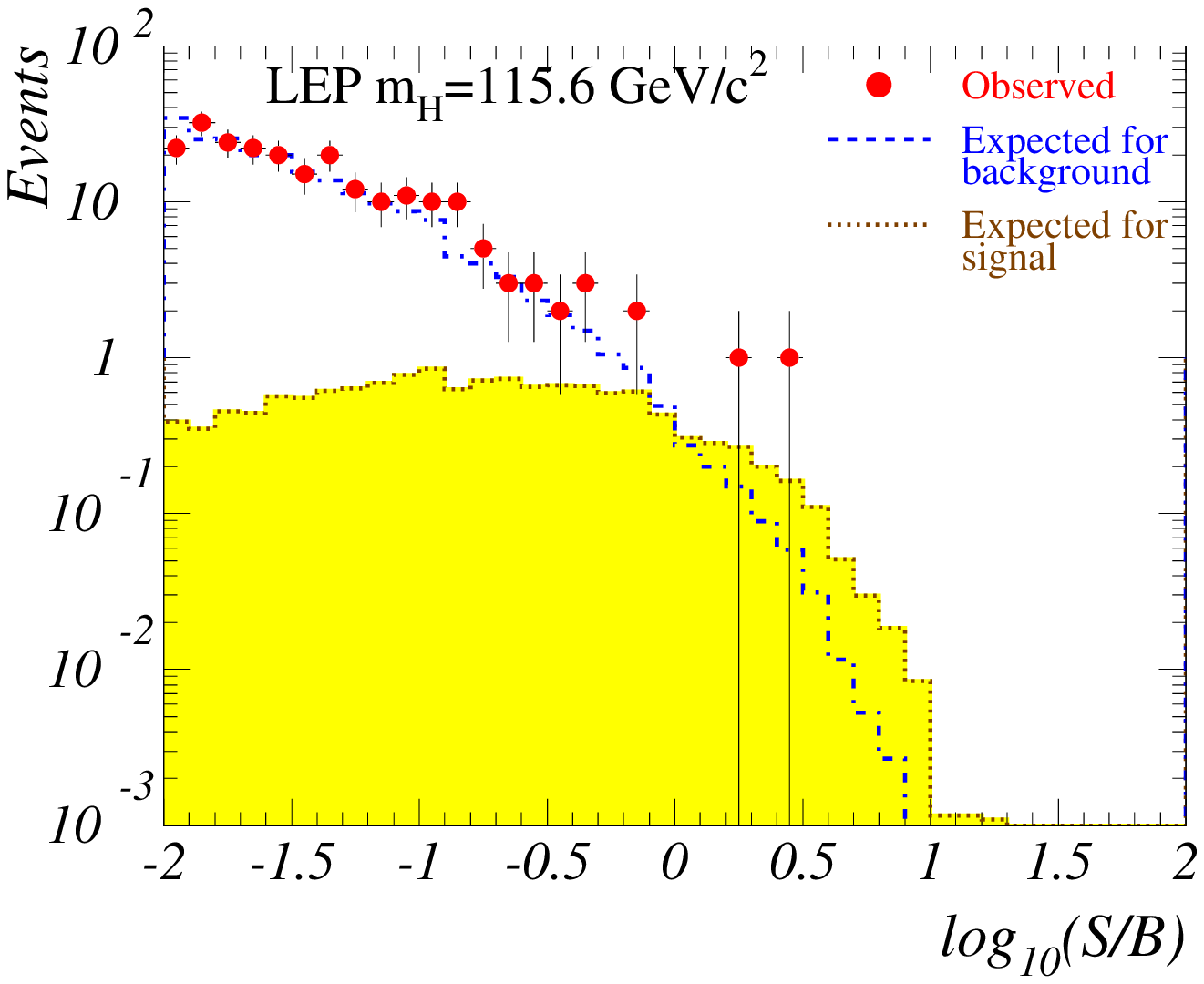,width=0.49\textwidth,height=6cm}
\epsfig{figure=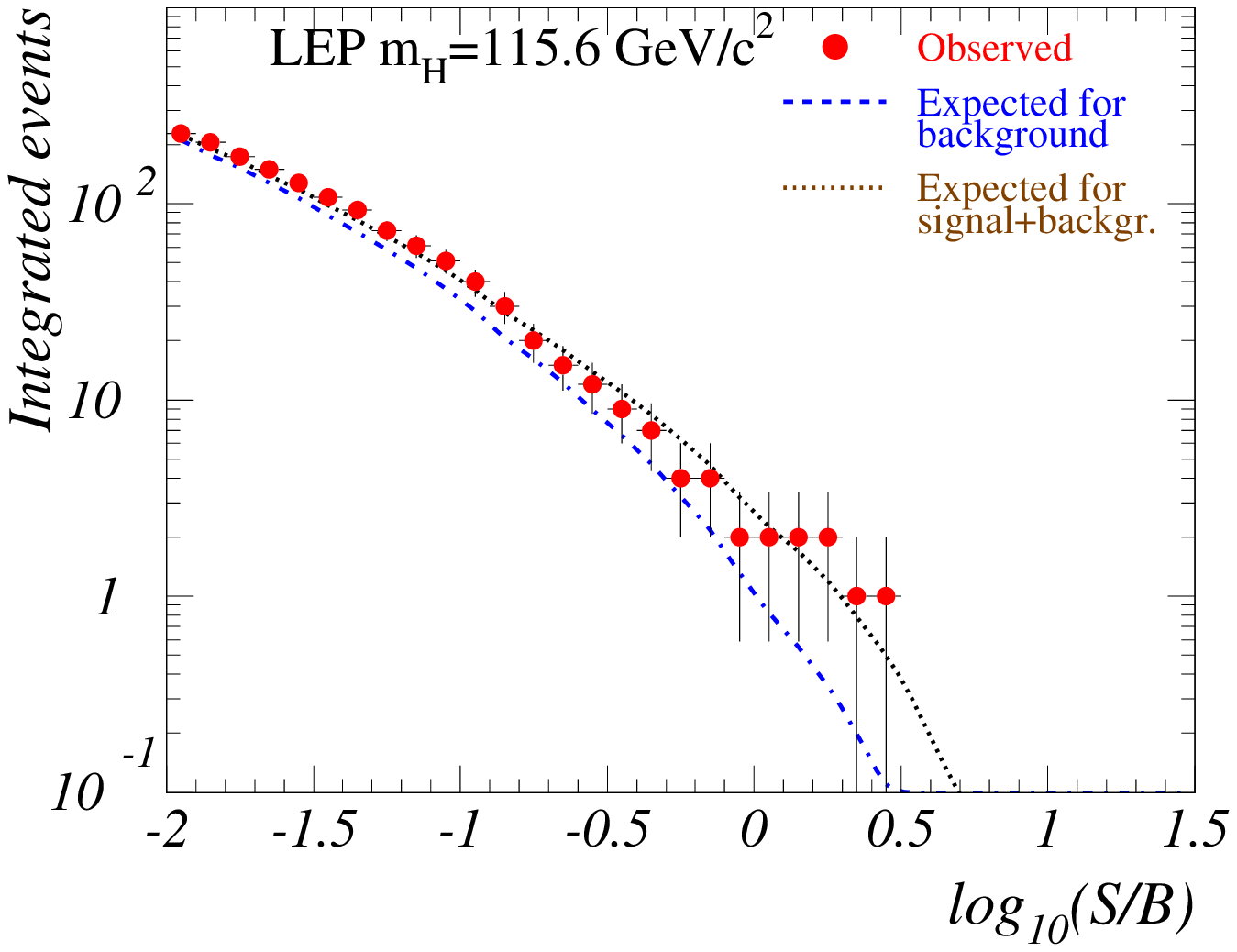,width=0.49\textwidth,height=6cm}\\
\epsfig{figure=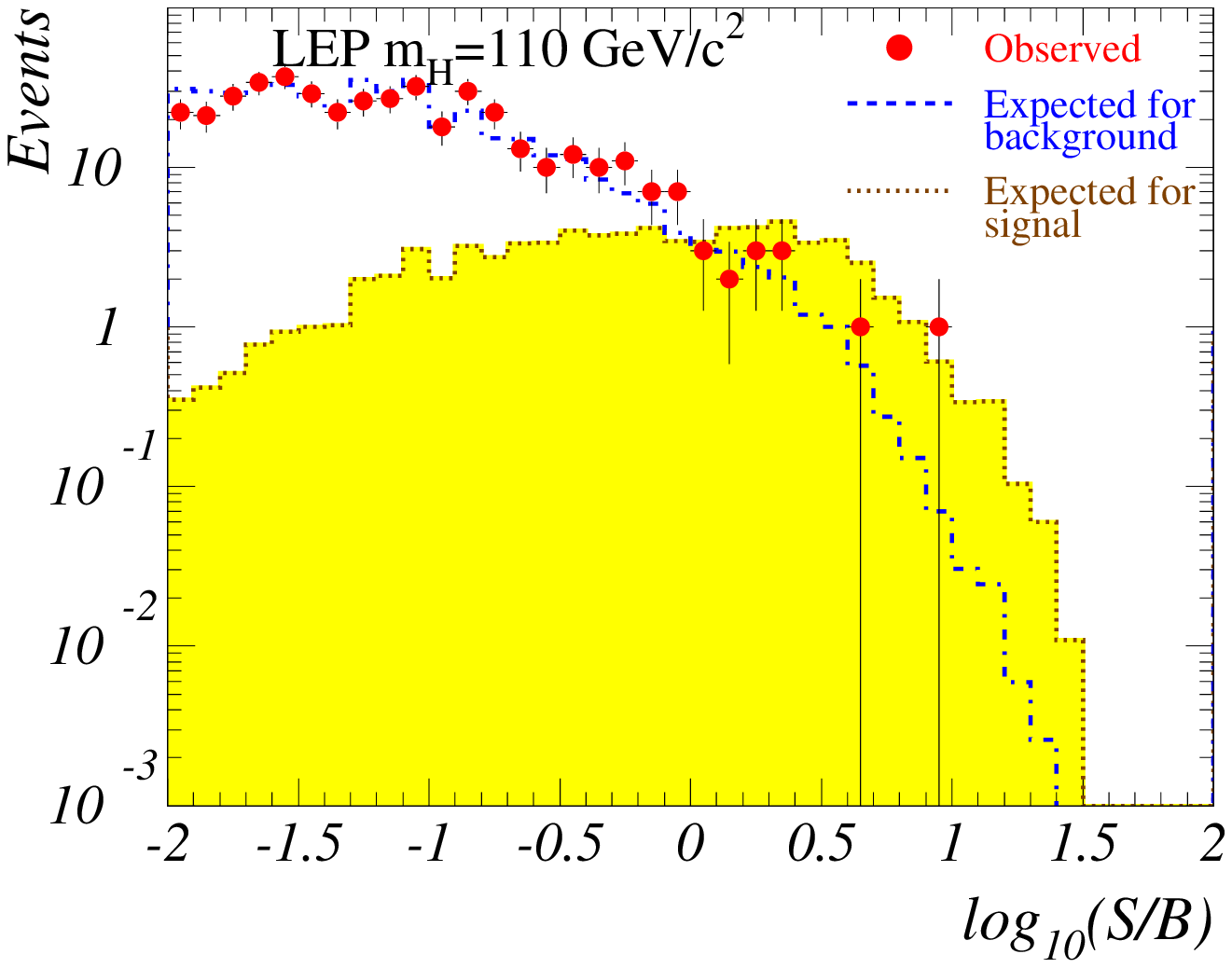,width=0.49\textwidth,height=6cm}
\epsfig{figure=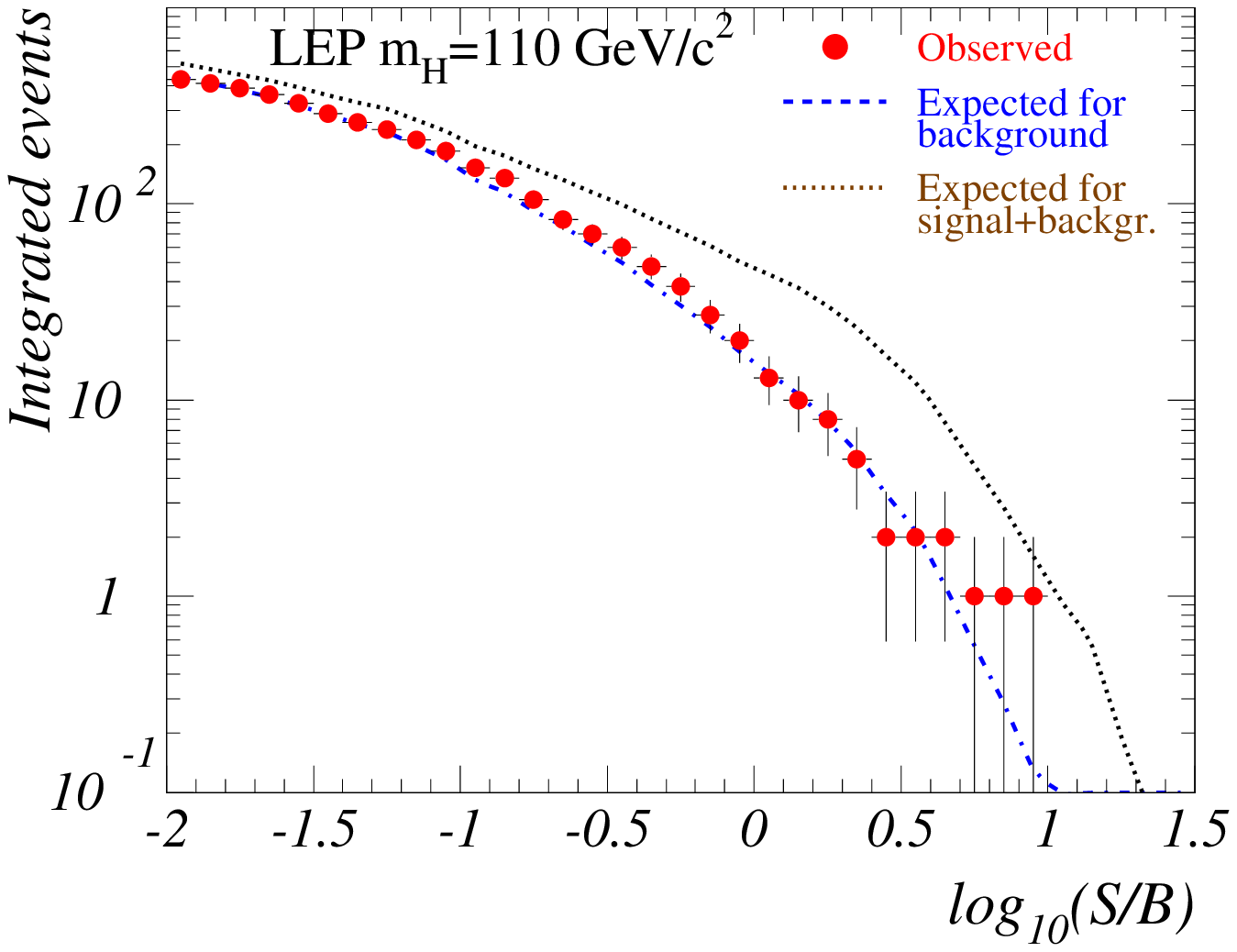,width=0.49\textwidth,height=6cm}\\
\caption[]{\small Left hand side: expected and observed distributions of $log_{10}s/b$ for 
a test-mass of $m_H=115.6$~GeV (upper part) 
and 110~GeV (lower part).
White/shaded histograms: expected distributions for the background/signal; points with error bars: selected data.
Right hand side: the integrals, from right to left, of the distributions shown in the plots on the left hand side. Dash-dotted/dotted
lines: expected for background/signal+background. 
\label{fig:adlo-weights}}
\end{center}
\end{figure}

The distribution of event weights for the test-mass fixed at $m_H=115.6$~GeV 
is shown in the upper part of Figure~\ref{fig:adlo-weights} ($\log_{10} s/b$ is plotted for better visibility). 
For the purpose of this figure, a cut at $s/b>0.01$ has been introduced.
The upper right plot shows the integrals of these distributions, starting from high values of $s/b$
(note that the bins are correlated).
The data prefer slightly the signal+background hypothesis over the background hypothesis although the separation is weak.
The two plots in the lower part show the corresponding distributions for a test-mass chosen arbitrarily 
at $m_H=110$~GeV. The data show clear preference for the background hypothesis in this case.

There is a general agreement between the observed and simulated rates, see Table~\ref{tab:rates}. 
\begin{table}[htb]
\begin{center}
\begin{tabular}{|c|c|cc|c|}
\hline
$m_H$       & $\ln(1+s/b)_{min}$  & Expected signal     & Expected backgd.     & Data   \\ 
\hline\hline     
            & $0.01$                & 75.7 & 440.0 & 430 \\
            & $0.1$                 & 63.2 & 128.7 & 151 \\
 110~GeV    & $0.18$                & 55.8 & 77.6  & 84  \\
            & $0.5$                 & 36.8 & 22.6  & 24  \\
            & $1$                   & 21.0 & 7.0   & 7   \\
            & $2$                   & 2.5  & 0.3   & 1   \\
\hline
            & $0.01$                & 17.7 & 226.5 & 242 \\
            & $0.1$                 & 11.4 & 34.8  & 47  \\
 115~GeV    & $0.18$                & 8.8  & 16.5  & 20  \\
            & $0.5$                 & 4.4  & 3.4   & 5   \\
            & $1$                   & 1.6  & 0.6   & 2   \\
            & $2$                   & 0.1  & 0.01  & 0   \\
\hline
            & $0.01$                & 13.4 & 211.7 & 227 \\
            & $0.1$                 & 7.7  & 26.2  & 38  \\
 115.6~GeV  & $0.18$                & 5.8  & 12.4  & 15  \\
            & $0.5$                 & 2.4  & 2.0   & 4   \\
            & $1$                   & 0.8  & 0.3   & 1   \\
            & $2$                   & 0.03 & 0.004 & 0   \\
\hline
\end{tabular}
\caption{\small Expected signal rates (for a SM Higgs boson with a mass of 110, 115, and 115.6~GeV) and
background rates, and the observed event count, for various cuts in $\ln(1+s/b)$.  
\label{tab:rates}}
\end{center}
\end{table}
\subsection{Distributions of the reconstructed Higgs boson mass}
\begin{figure}[htb]
\begin{center}
\epsfig{figure=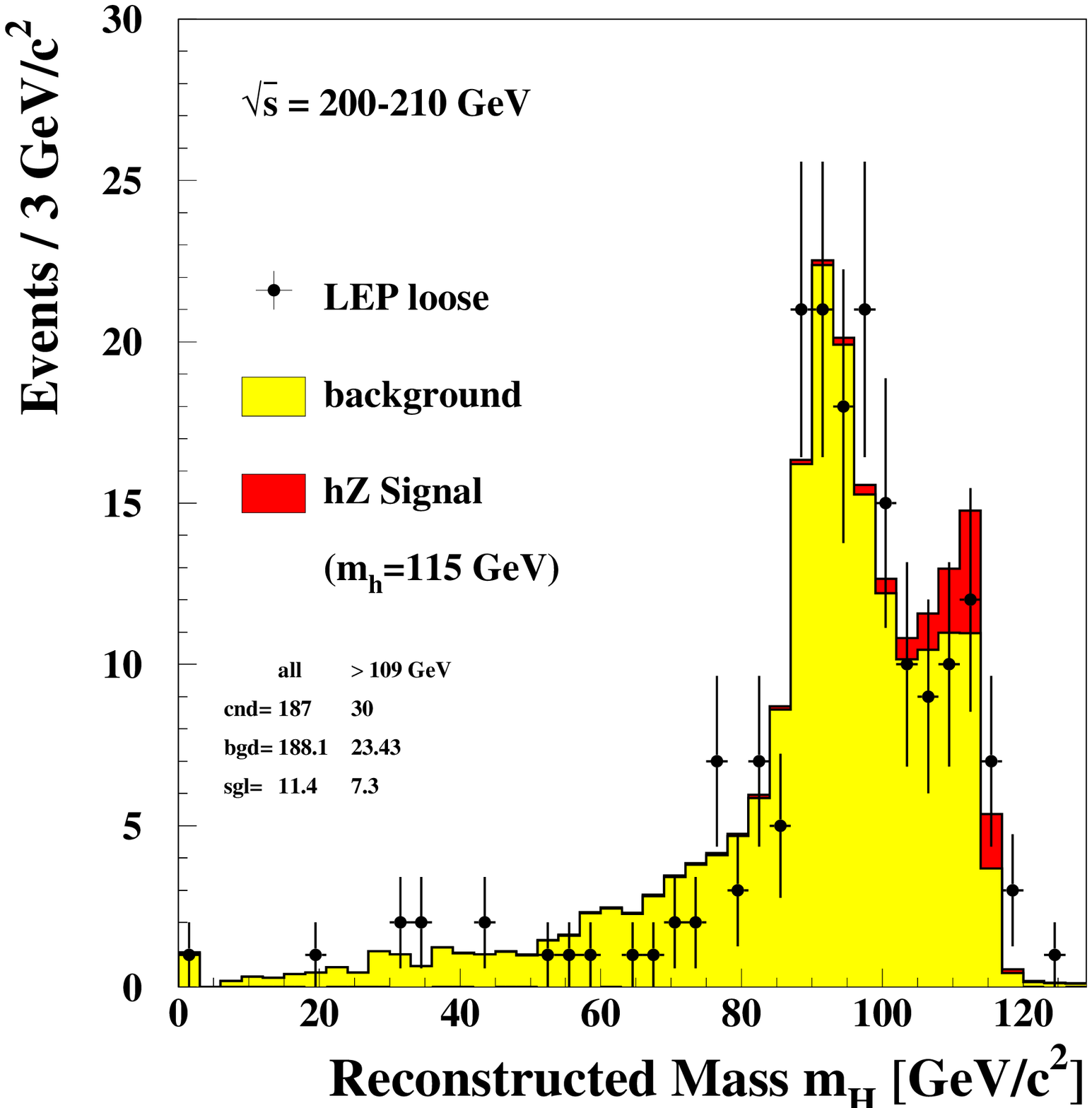,width=0.49\textwidth}
\epsfig{figure=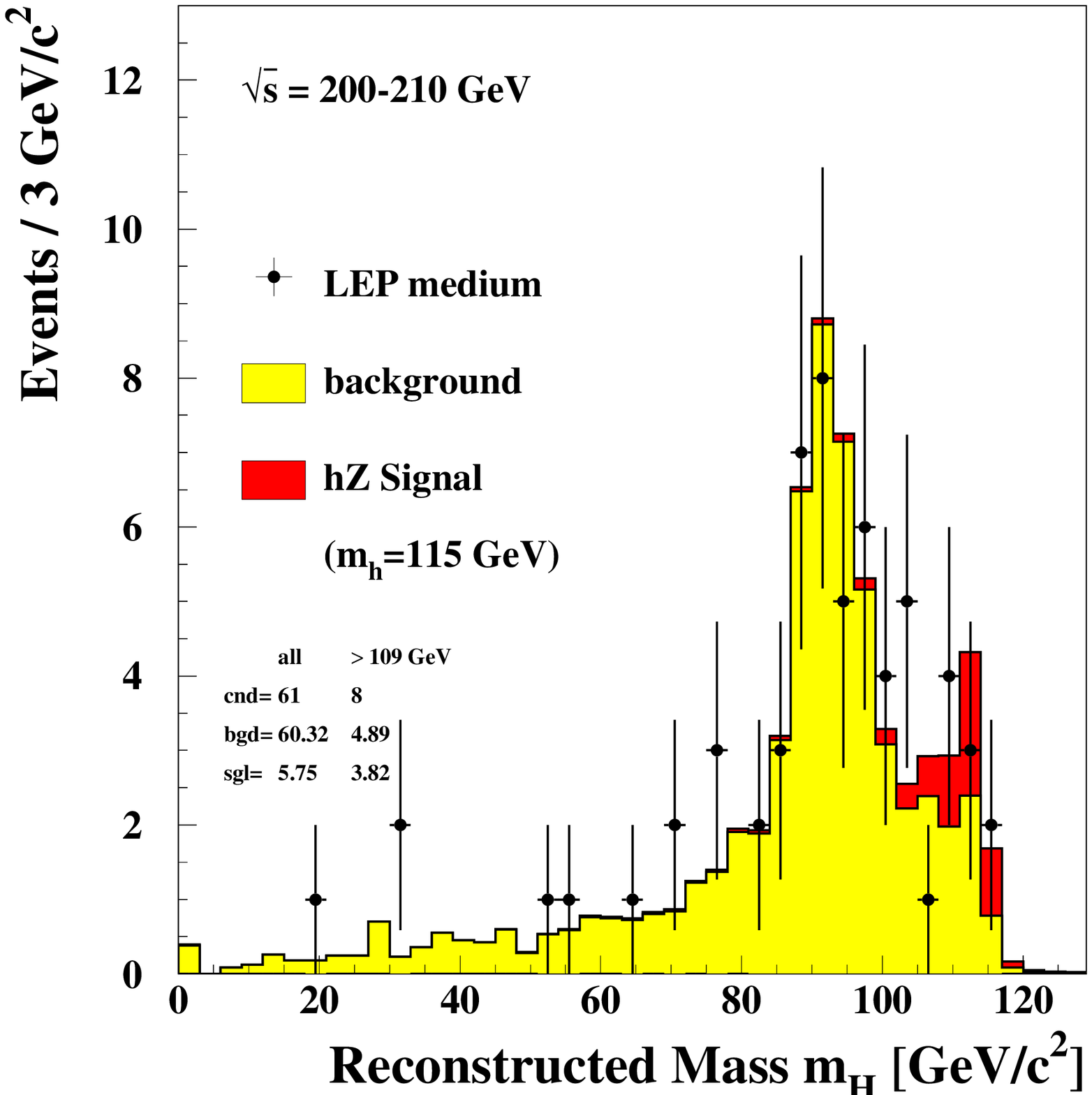,width=0.49\textwidth}\\
\begin{center}
\epsfig{figure=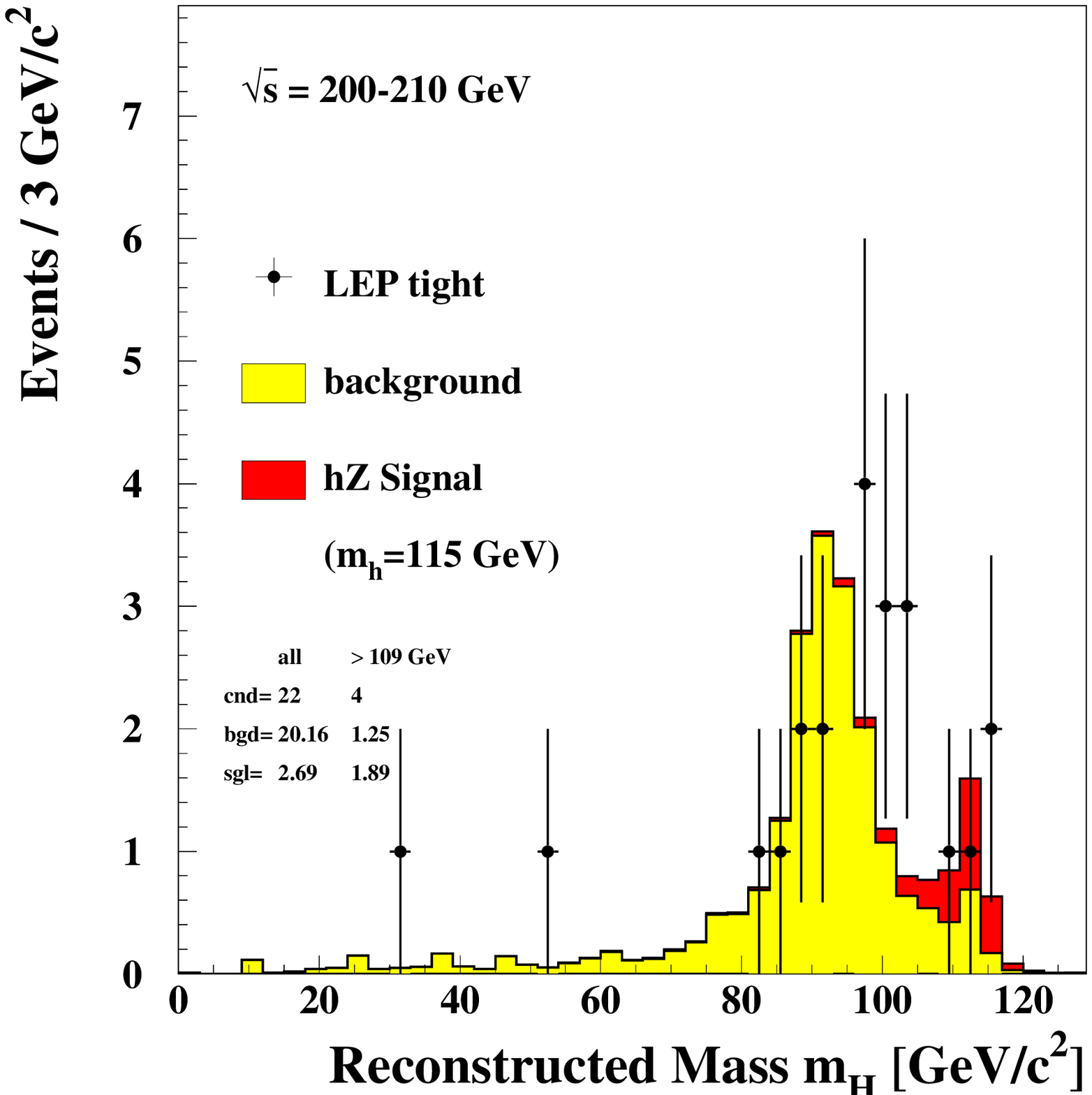,width=0.49\textwidth}\\
\end{center}
\caption[]{\small Distributions of the reconstructed Higgs mass, $m_H^{rec}$, 
from three special, non-biasing, selections with increasing purity of a signal from a 115 GeV Higgs boson.
\label{fig:masses}}
\end{center}
\end{figure}

The reconstructed Higgs boson mass $m_H^{rec}$ is just one of several discriminating variables contributing to the
separation of the signal and the background processes and the construction of the likelihood ratio $Q$. 
Since in some channels the event selection depends explicitly
on the test-mass, the reconstructed mass distributions resulting from the standard combination procedure are biased. 
The distributions shown in Figure~\ref{fig:masses}
are therefore obtained from special selections where the cuts are applied on quantities (e.g. b-tag variables) 
which introduce little bias into the $m^{rec}_H$ distribution. Three such selections are shown, with increasing signal purity.
In the loose/medium/tight selections the cuts are adjusted in each decay channel to obtain for $m_H=115~GeV$ a signal over background 
ratio
\footnote{The signal-to background ratio used in these
selection is different from the ratio $s/b$ describing event weights.}
of 0.5/1/2 in the reconstructed mass region 
above 109~GeV.  
These spectra are shown merely to illustrate the agreement between the data and the simulation 
in this important discriminating variable, and should not be used to draw conclusions regarding the 
significance of a possible signal. Most importantly, it is not claimed that the slight excess at high mass 
in the tight selection (4 events for an expected background of 1.25 events) is solely responsible for the result quoted below.
\subsection{Confidence level calculation}
\begin{figure}[htb]
\begin{center}
\epsfig{figure=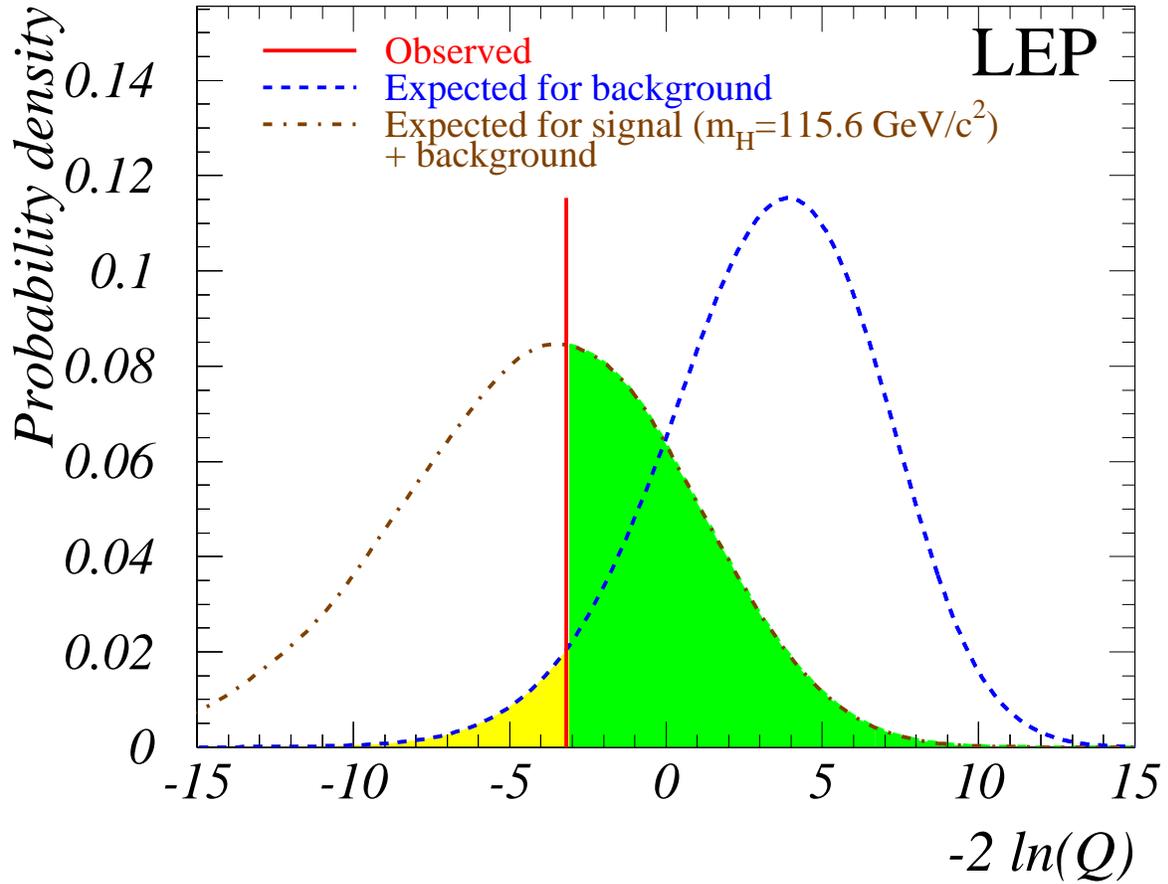,width=\textwidth}\\
\caption[]{\small Probability density functions corresponding to a test-mass $m_H=115.6$~GeV, for the background 
and signal+background hypotheses. The observed value of $-2\ln Q$ which corresponds to the data is indicated by the 
vertical line. The light shaded region is a measure of the compatibility with the background hypothesis, $1-CL_b$,
and the dark shaded region is a measure of compatibility with the signal+background hypothesis, $CL_{s+b}$.
\label{fig:adlo-prob-dens}}
\end{center}
\end{figure}

The expected distributions of $-2\ln Q$ for a test-mass of 115.6~GeV (a slice of Figure~\ref{fig:adlo-lnq} 
at $m_H=115.6$~GeV)
are shown in Figure~\ref{fig:adlo-prob-dens}. The distributions for the background and the signal+background hypotheses
are normalized and represent probability density functions. The vertical line indicating the observed value lies within the
distribution for the signal+background hypothesis. 
The integral of the background distribution from $-\infty$ to the observed value, $1-CL_b$, measures the 
compatibility of the observation with the background hypothesis. Given a large number of background experiments, 
it is the probability to obtain an event configuration more signal-like than the one observed.
Similarly, the integral from  $+\infty$ to the observed value of the signal+background distribution,
$CL_{s+b}$, is a measure of compatibility with the signal+background hypothesis.  

\begin{figure}[htb]
\begin{center}
\epsfig{figure=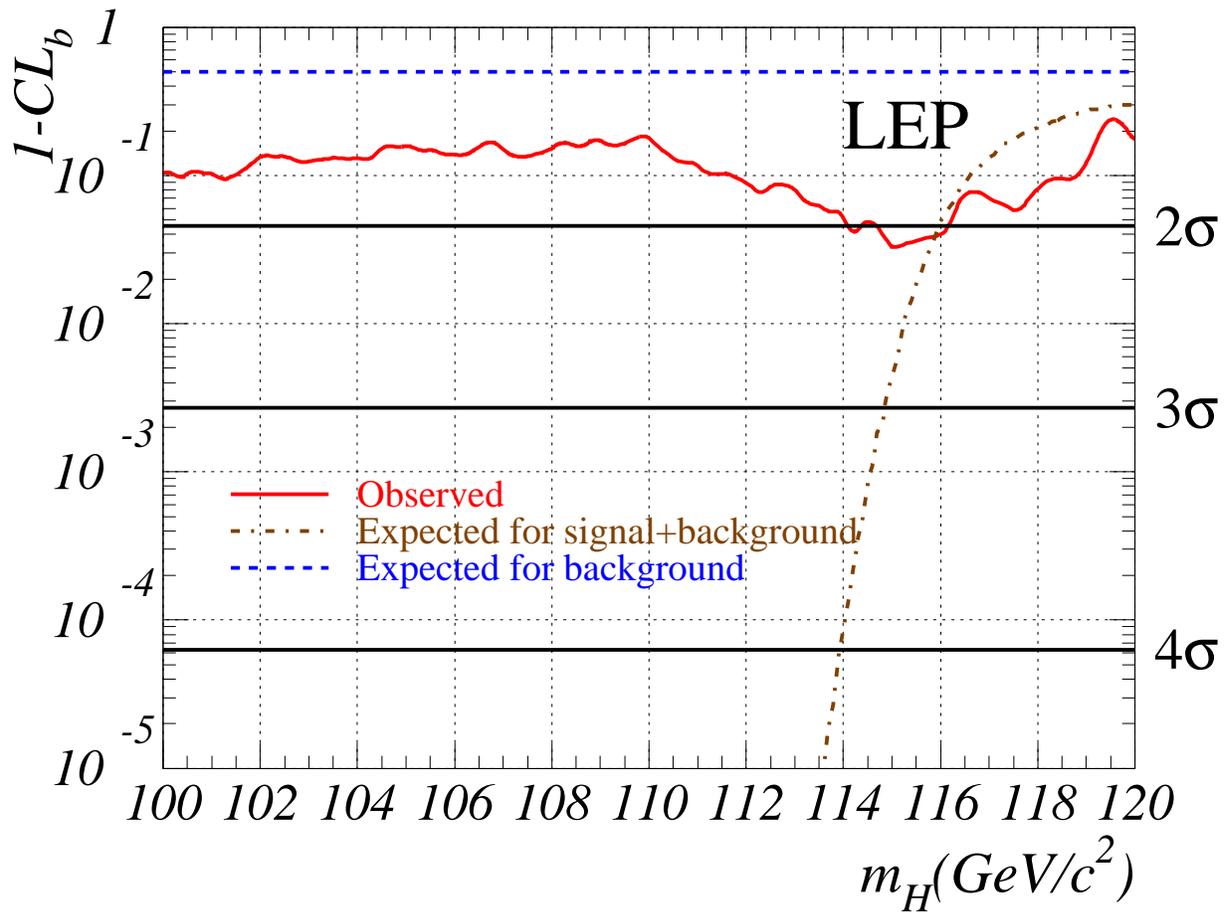,width=\textwidth}\\
\caption[]{\small The probability $1-CL_b$ as a function of the test-mass $m_H$. Solid line: observation; 
dashed/dash-dotted lines: expected probability for the background/signal+background hypotheses. 
See Footnote $^4$ for the transformation of $1-CL_b$ values into standard deviations.
\label{fig:adlo-clb}}
\end{center}
\end{figure}

Calculating $1-CL_b$ for test-masses between 100 and 120~GeV, Figure~\ref{fig:adlo-clb} is obtained. 
At $m_H=115.6$~GeV, where the $-2\ln Q$ has its minimum (see Figure~\ref{fig:adlo-lnq}), 
one gets $1-CL_b = 0.034$, which corresponds to about two standard deviations
\footnote{For the conversion of $1-CL_b$ into 
standard deviations $(\sigma)$, we adopt  
a gaussian approximation~\cite{pdg-stat} and use a ``one-sided" convention where 
$1-CL_b = 2.7\times 10^{-3}$ would indicate a $3\sigma$ ``evidence" and $1-CL_b = 5.7\times 10^{-7}$ 
a $5\sigma$ ``discovery". The median expectation for pure background is 0.5. Values smaller or larger than 0.5 indicate 
an excess or deficit, respectively. In this scheme, the current result, $1-CL_b=0.034$, corresponds to 2.1$\sigma$. 
The earlier LEP results quoted in
the first and second line of Table~\ref{tab:history} correspond to $2.2\sigma$ and $2.9\sigma$, respectively.
This convention is also used in Figure~\ref{fig:adlo-clb} to indicate the levels of significance on the right-hand 
scale.
The $\pm 1$ and $\pm 2$ standard deviation ``bands" which show up e.g. in the $-2\ln Q$ plots correspond to a
slightly different, ``two-sided", convention.
}. 
Values of $1-CL_b$ and $CL_s$ corresponding to $\mH=115.6$~GeV are
listed in Table~\ref{tab:clb}. 
\begin{table}[htb]
\begin{center}
\begin{tabular}{|l|cc|}
\hline
           & $1-CL_b$              & $CL_{s+b}$        \\
\hline\hline
ALEPH      &  $2.0\times 10^{-3}$  &  0.94             \\
DELPHI     &  0.87                 &  0.02             \\
L3         &  0.24                 &  0.47             \\
OPAL       &  0.22                 &  0.47             \\
\hline
DLO        &  0.49                 &  0.07             \\
ALO        &  $3.7\times 10^{-3}$  &  0.83             \\
\hline
Four-jet   &  0.016                &  0.74             \\
Missing energy  &  0.40            &  0.26             \\
\hline
All but four-jet & 0.34            & 0.19              \\ 
\hline\hline
LEP       &   0.034                &  0.44             \\
\hline
\end{tabular}
\caption{\small The background probability $1-CL_b$  and the signal+background 
probability $CL_{s+b}$ at $m_H=115.6$~GeV, 
for subsets and for all LEP data. 
DLO/ALO designate subset where the ALEPH/DELPHI data are left out of the combination.
\label{tab:clb}}
\end{center}
\end{table}

It should be noted that these probabilities refer to {\it local} fluctuations of the background.
To obtain the probability for such a fluctuation to appear anywhere within a given mass range of interest,
a multiplicative factor has to be applied which is approximated by the width of the mass range divided by the mass
resolution. In the present case the range of interest is limited from below by previous exclusion limits 
(107.9~GeV~\cite{adlo-cernep}) and from above
by the kinematic limit of the production process \ee\ra~HZ (about 116~GeV). The mass resolution 
averaged over the final-state topologies and experiments is about 3.5~GeV.

%
\subsection{Bounds for the Higgs boson mass and coupling}
\begin{figure}[htb]
\begin{center}
\epsfig{figure=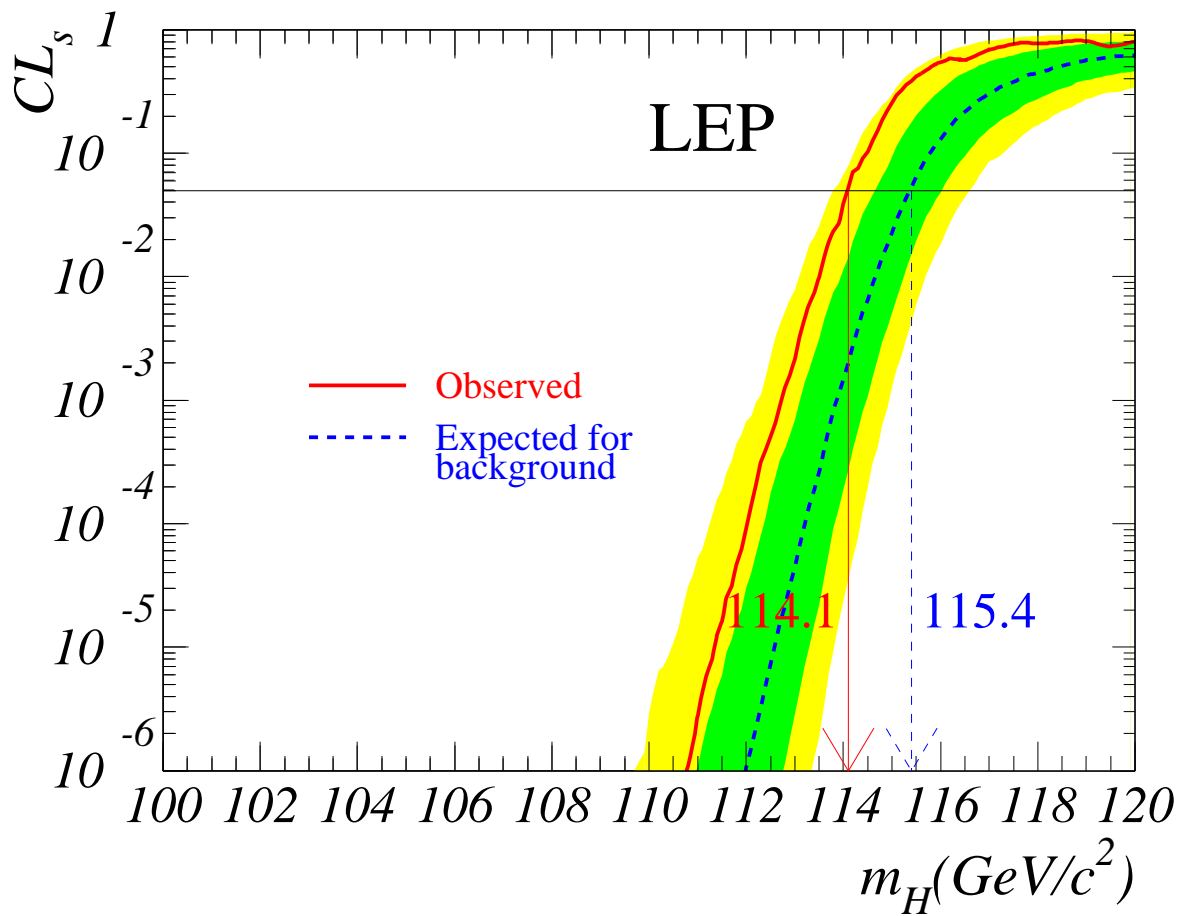,width=\textwidth}
\caption[]{\small Confidence level $CL_s$ for the signal+background hypothesis. Solid line: observation; dashed line:
median background expectation. The dark/light shaded bands around the median
expected line correspond to the $\pm 1$/$\pm 2$ standard deviation spreads from a large number of background experiments.
\label{fig:adlo-cls}}
\end{center}
\end{figure}

The ratio $CL_s = CL_{s+b}/CL_b$ as a function of the test-mass, shown in 
Figure~\ref{fig:adlo-cls}, is used to derive a lower bound for the SM Higgs boson mass~(\cite{adlo-cernep}, Appendix A). 
The test-mass corresponding to $CL_s=0.05$   
defines the lower bound at the 95\% confidence level.
\begin{table}[htb]
\begin{center}
\begin{tabular}{|l|cc|}
\hline
                               & Expected limit (GeV) & Observed limit (GeV)    \\
\hline\hline
ALEPH                          &    113.8             &   111.5   \\   
DELPHI                         &    113.5             &   114.3   \\ 
L3                             &    112.7             &   112.2   \\ 
OPAL                           &    112.6             &   109.4   \\
\hline
DLO                            &    114.9             &   114.8   \\ 
\hline\hline
LEP                            &    115.4             &   114.1   \\
\hline
\end{tabular}
\caption{\small Expected (median)  and observed 95\% CL lower bounds on the SM Higgs boson mass, 
for the individual experiments, for DLO (with ALEPH left out of the combination) and for all LEP data combined.
\label{tab:mass-limits}}
\end{center}
\end{table}
The expected and observed lower bounds obtained for the SM Higgs boson mass are listed in Table~\ref{tab:mass-limits}.
The current lower bound from LEP is 114.1~GeV at the 95\% confidence level.

\begin{figure}[htb]
\begin{center}
\epsfig{figure=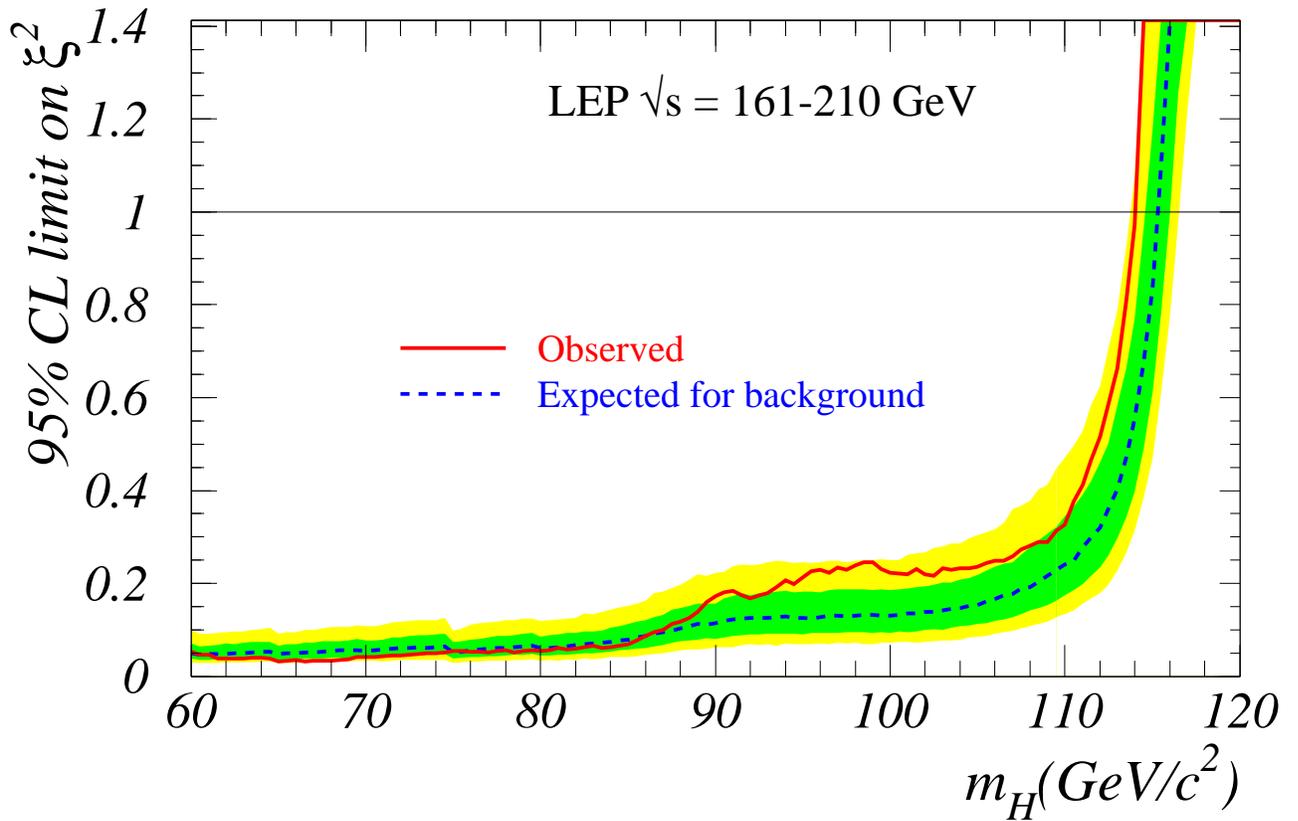,width=\textwidth} 
\caption[]{\small The 95\% CL upper bound on $\xi^2$ as a function 
of \mH, where $\xi= g_{HZZ}/g_{HZZ}^{SM}$ is the HZZ coupling relative to the SM
coupling. The dark/light shaded bands around the median
expected line correspond to the $\pm 1$/$\pm 2$ standard deviation spreads from a large number 
of background experiments. The horizontal line corresponds to the SM coupling.
\label{sm-xi2}}
\end{center}
\end{figure}

The LEP data are used also to 
set 95\% CL upper bounds
on the square of the HZZ coupling in non-standard models which assume the same Higgs
decay properties as in the SM but where 
the HZZ coupling
may be different. Figure~\ref{sm-xi2} shows the upper bound on 
$\xi^2= (g_{HZZ}/g_{HZZ}^{SM})^2$, the square of the ratio of the coupling in such a model to the SM coupling, 
as a function of the Higgs boson mass. In deriving this limit, the data collected at $E_{cm}$ = 161, 172 and 183~GeV were also
included. 
%
%
\section{Cross-checks, uncertainties}
\begin{figure}[htb]
\begin{center}
\epsfig{figure=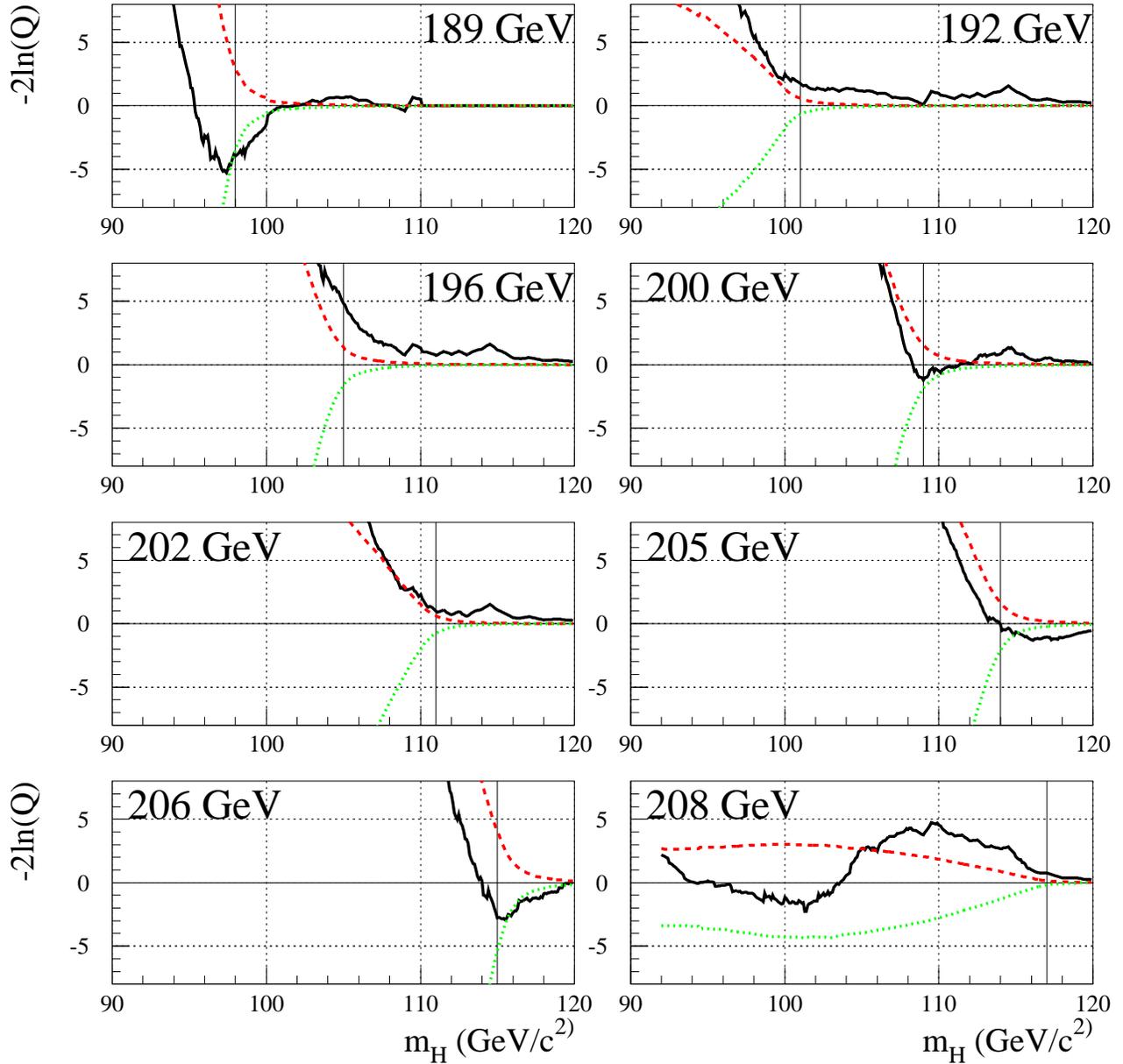,width=\textwidth}
\caption[]{\small Behaviour of $-2\ln Q$ in subsets collected at different c.m. energies.
In each plot, the full curve shows the observed behaviour, the dashed/dotted
lines show the expected behaviour for background/signal+background, and the vertical line indicates the 
test-mass $m = E_{cm} - M_Z$~GeV, just at the kinematic limit. (The subset labelled 208 GeV has very low
statistics.)
\label{fig:threshold}}
\end{center}
\end{figure}

(i) It is legitimate to ask whether the excess at 115~GeV mass could be induced by an inadequate treatment of
the data close to the kinematic limit of the process \ee\ra~\Ho\Z. To test this hypothesis, 
the $-2\ln Q$ curves (the equivalents to Figure~\ref{fig:adlo-lnq}) have been produced separately for 
data of different c.m. energies, see Figure~\ref{fig:threshold} . In each plot, 
the vertical line indicates the test-mass $m_H = E_{cm} - M_Z$~GeV, just at the kinematic limit. 
%
%

In the 189~GeV data, an excess at $m=97$~GeV has indeed been observed~\cite{adlo-1999} (see the large negative
value of $-2\ln Q$ close to the signal+background prediction) which
was due mainly to small excesses in ALEPH and OPAL data compatible with \ee\ra~\Z\Z,
the dominant background in the vicinity of that mass. This excess still has
a significance of about two standard deviations when LEP data from all energies are combined, 
and one cannot exclude a physics interpretation beyond the SM (e.g. MSSM with several neutral Higgs bosons). 
However, there is no evidence for a systematic effect at threshold in the data collected at the other energies below 206~GeV. 

(ii) The LEP experiments claim systematic errors of typically 5\% for their signal estimates and 10\% for their 
background estimates. Most of the errors are estimated
from calibration data (e.g. data taken at $E_{cm}=M_Z$ to calibrate the b-tagging performance 
or to determine the level of non-b background) or from measurements of the \ee\ annihilations into fermion pairs,
WW and ZZ processes.
The current implementation of systematic uncertainties (see Ref's~\cite{aleph}-\cite{opal} for details)
treats errors from the same source as fully correlated between experiments and errors from different 
sources as uncorrelated.
Furthermore, all bins within the same channel have the same errors, and these 
errors are assumed to have Gaussian distributions. 
Several tests have been performed to assess the possible impact of this simplified treatment on the result.
 
(a) If the systematic errors are ignored, $1-CL_b$ decreases from 3.4\% to 3.2\%.

(b) The backgrounds in all channels have to be increased coherently 
by 13\% to reduce the excess at 115.6~GeV 
to the level of one standard deviation, and by 26\% to get a typical background result ($1-CL_b=0.5$). 
Such large coherent changes are not consistent with the quoted error estimates.

\begin{figure}[htb]
\begin{center}
\epsfig{figure=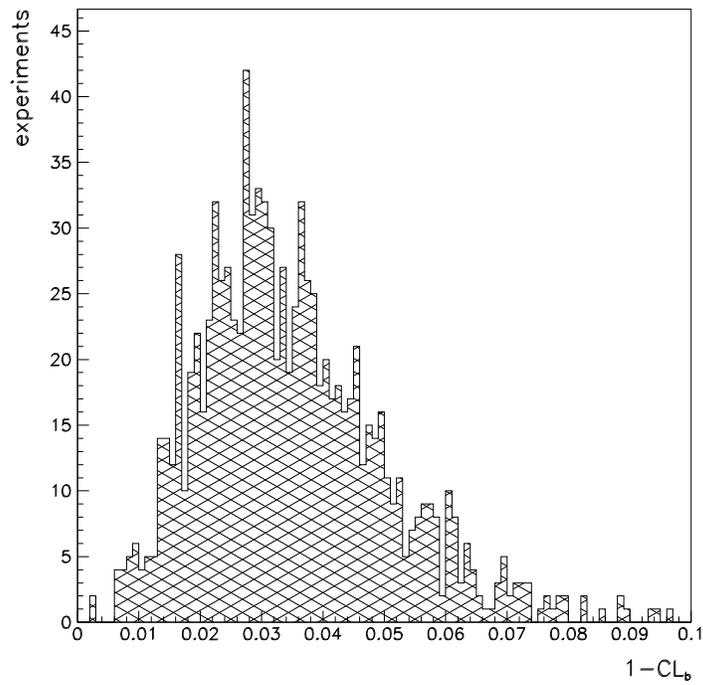,width=0.6\textwidth}\\
\caption[]{\small Distribution of the background probability $1-CL_b$ for a test-mass of 115.6~GeV 
obtained from 1000 simulated
experiments where the expected signal and background has been varied randomly according to the systematic
errors and their correlations.
\label{fig:bock-error}}
\end{center}
\end{figure}

(c) In a test, the value of $1-CL_b$ for the observed data 
was recomputed 1000 times, each time with a set of signal and background
estimations chosen randomly according to the assigned systematic uncertainties and their correlations.
The distribution of $1-CL_b$ at $m_H=115.6$~GeV is shown in Figure~\ref{fig:bock-error}. 
From the r.m.s. width of the distribution (about 50\% of the mean value) and its asymmetry, one can
conclude that the spread of results one can obtain by varying the signal and background levels according to their
errors is of approximately $\pm 0.2$ standard deviations, when $1-CL_b$ is interpreted in terms
of standard deviations.   The systematic errors are already incorporated into the quoted
result, and this information is provided to demonstrate the limited sensitivity to the quoted systematic effects.

These tests do not address the question of completeness of the systematic errors provided by the experiments 
for the combination. Since only one of the experimental collaborations has published its final results,
changes to the systematic errors provided by the other experiments cannot be excluded. 

(iii) A technical uncertainty is ascribed to various approximations 
which are necessary to speed up the computations. This uncertainty 
is estimated by comparing the results from different software packages and by 
reproducing the $-2\ln Q$, $1-CL_b$ and $CL_s$ results of individual experiments prior to the combination. 
For the present paper, the value of $1-CL_b$ in the vicinity of $m_H=115.6$~GeV has been determined 
independently by four combiners; they fall within a range of $\pm 5$\% (relative). The highest value, 
$1-CL_b=0.034$, is retained as the result.
\section{Internal consistency}
The excess at $m_H=115.6$~GeV has been examined in subsets obtained by dividing the data 
by experiment and by decay channel. It has also been analysed as a function of signal purity. 

\begin{figure}[htb]
\begin{center}
\epsfig{figure=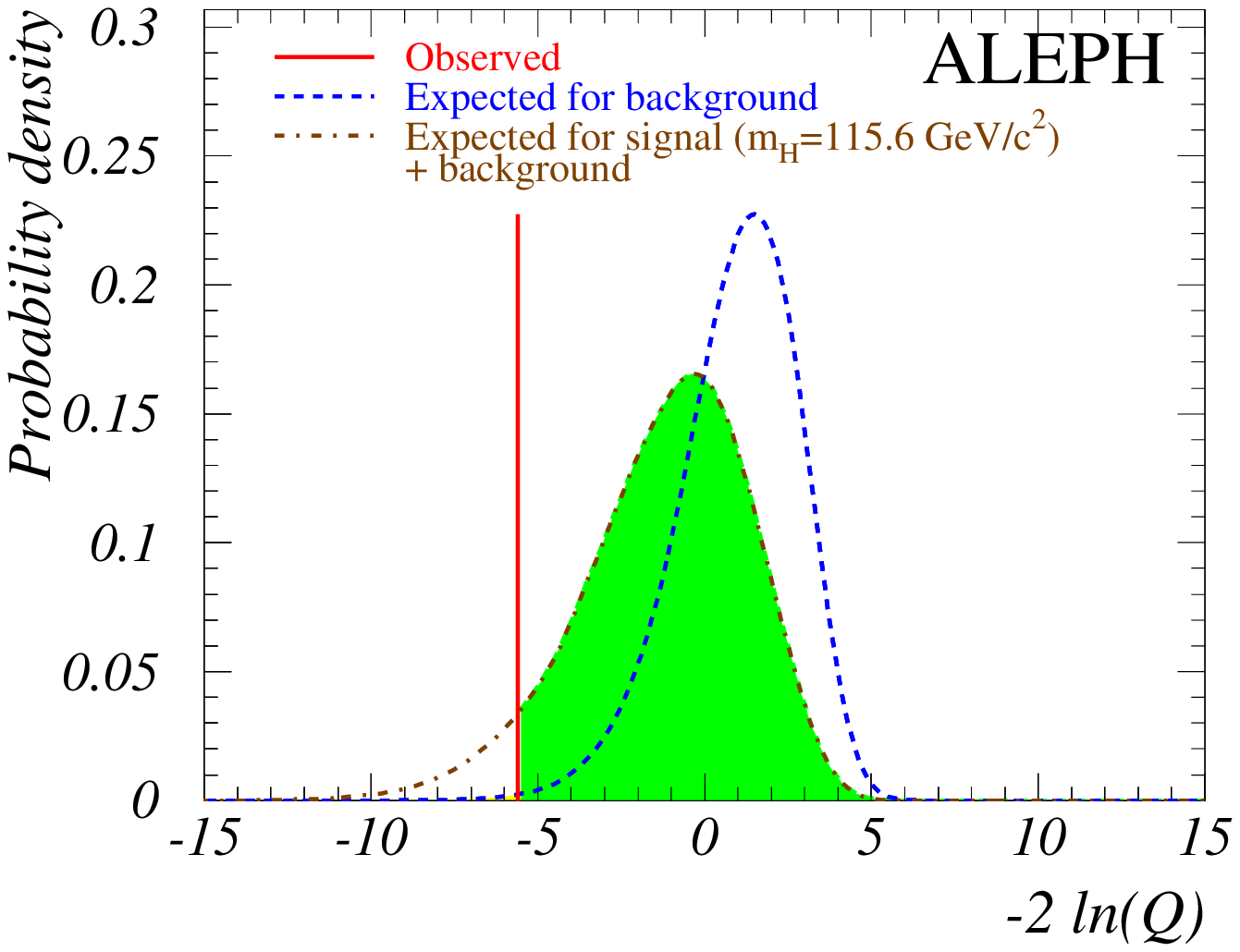,width=0.49\textwidth,height=4cm}
\epsfig{figure=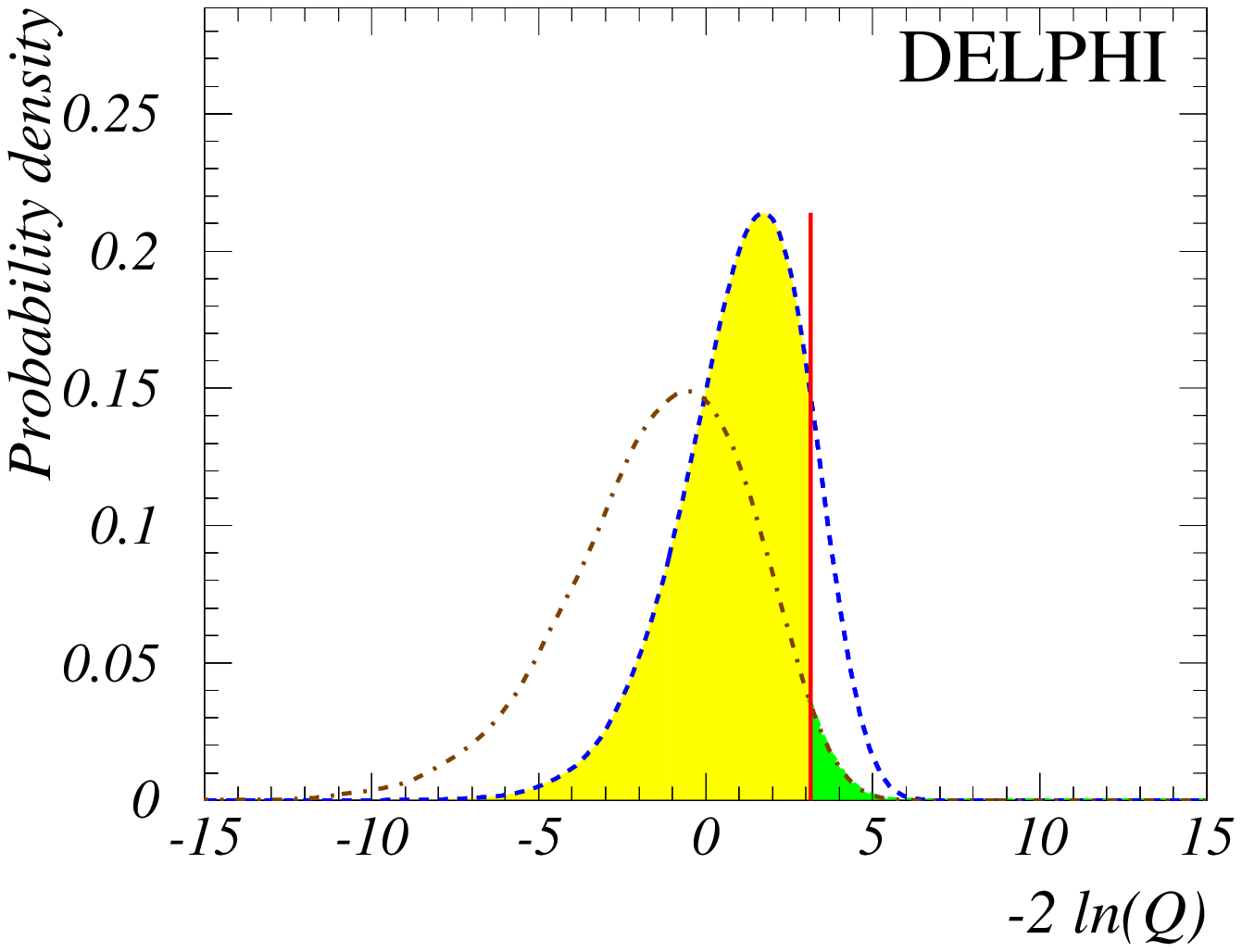,width=0.49\textwidth,height=4cm}\\
\epsfig{figure=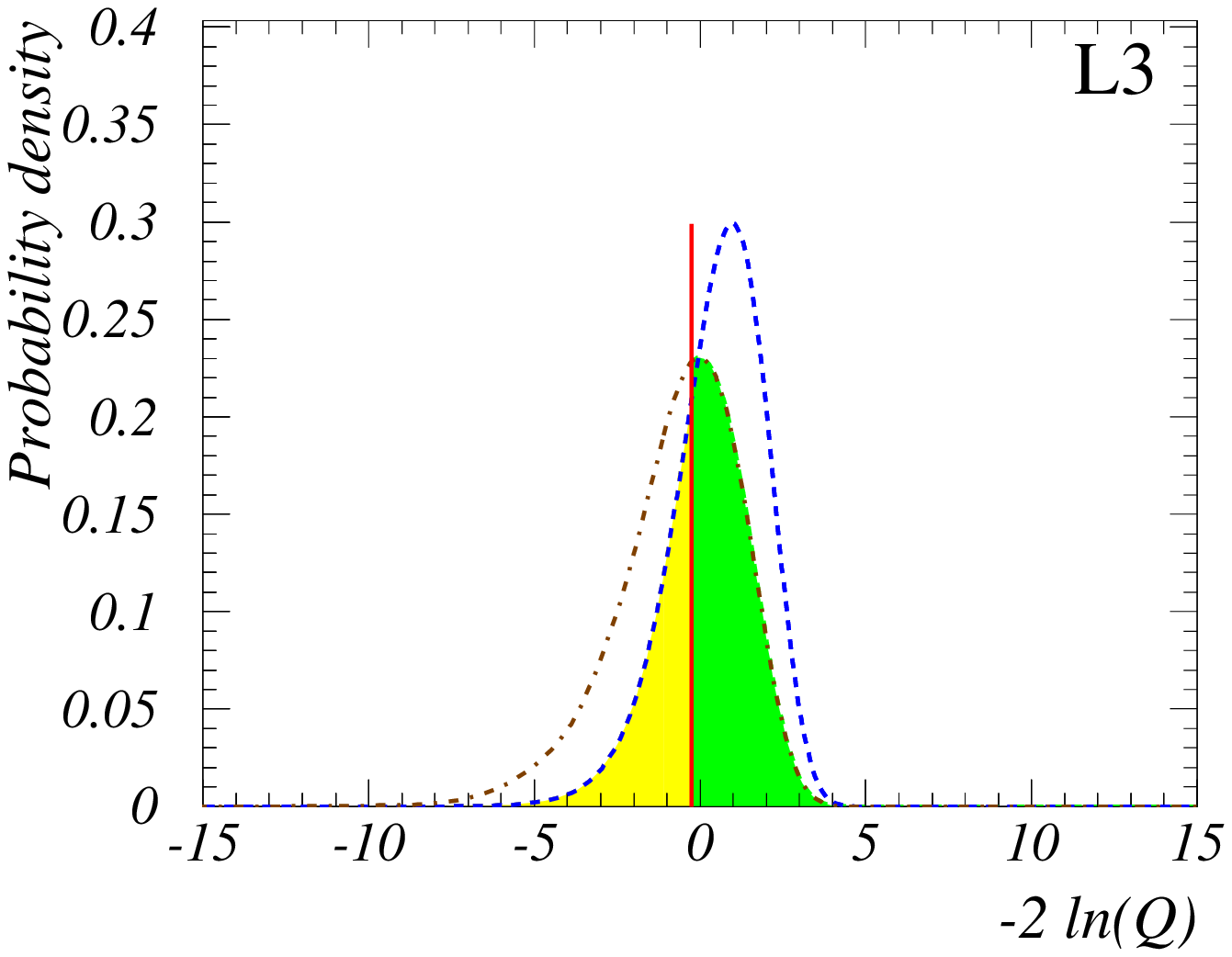,width=0.49\textwidth,height=4cm}
\epsfig{figure=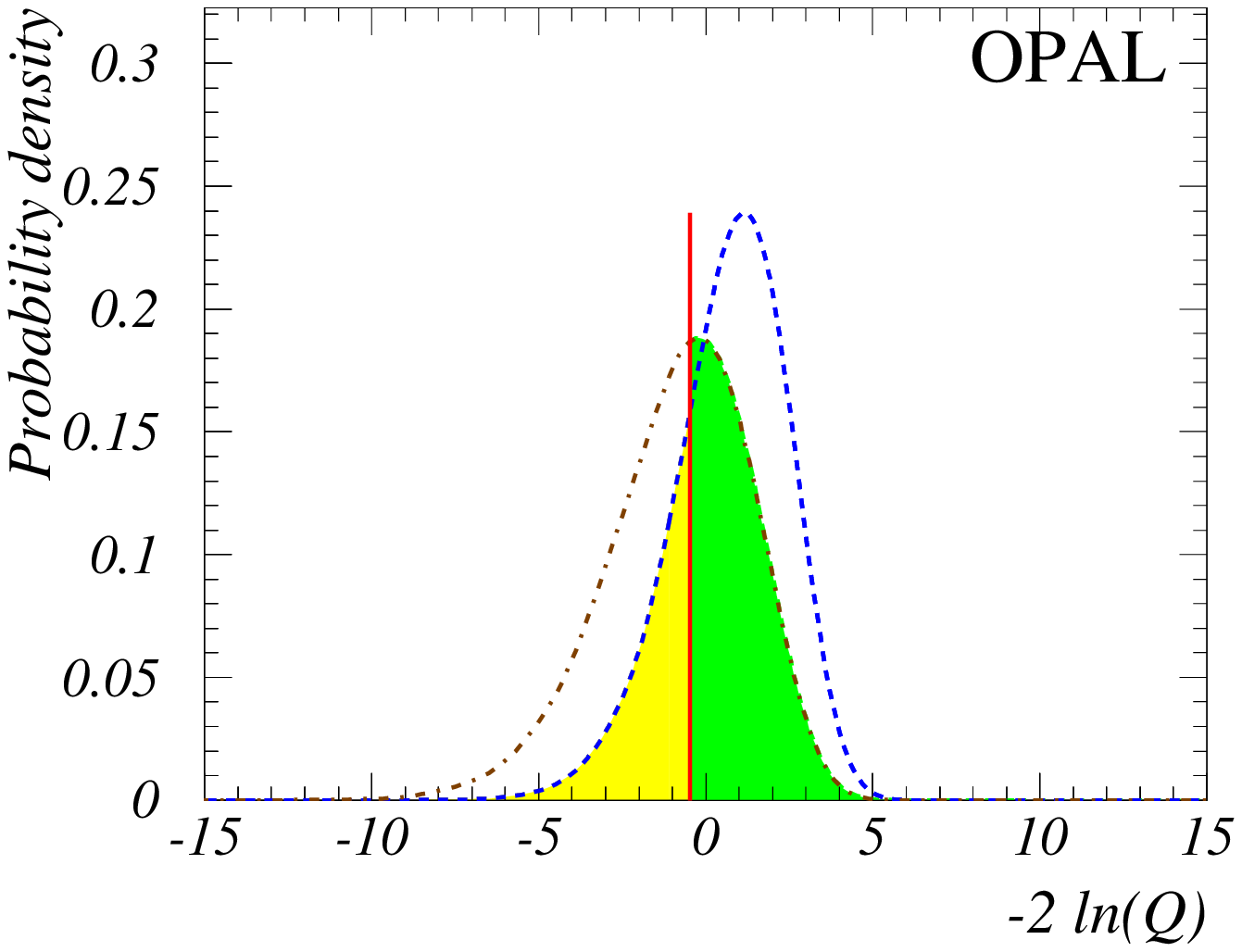,width=0.49\textwidth,height=4cm}\\
\epsfig{figure=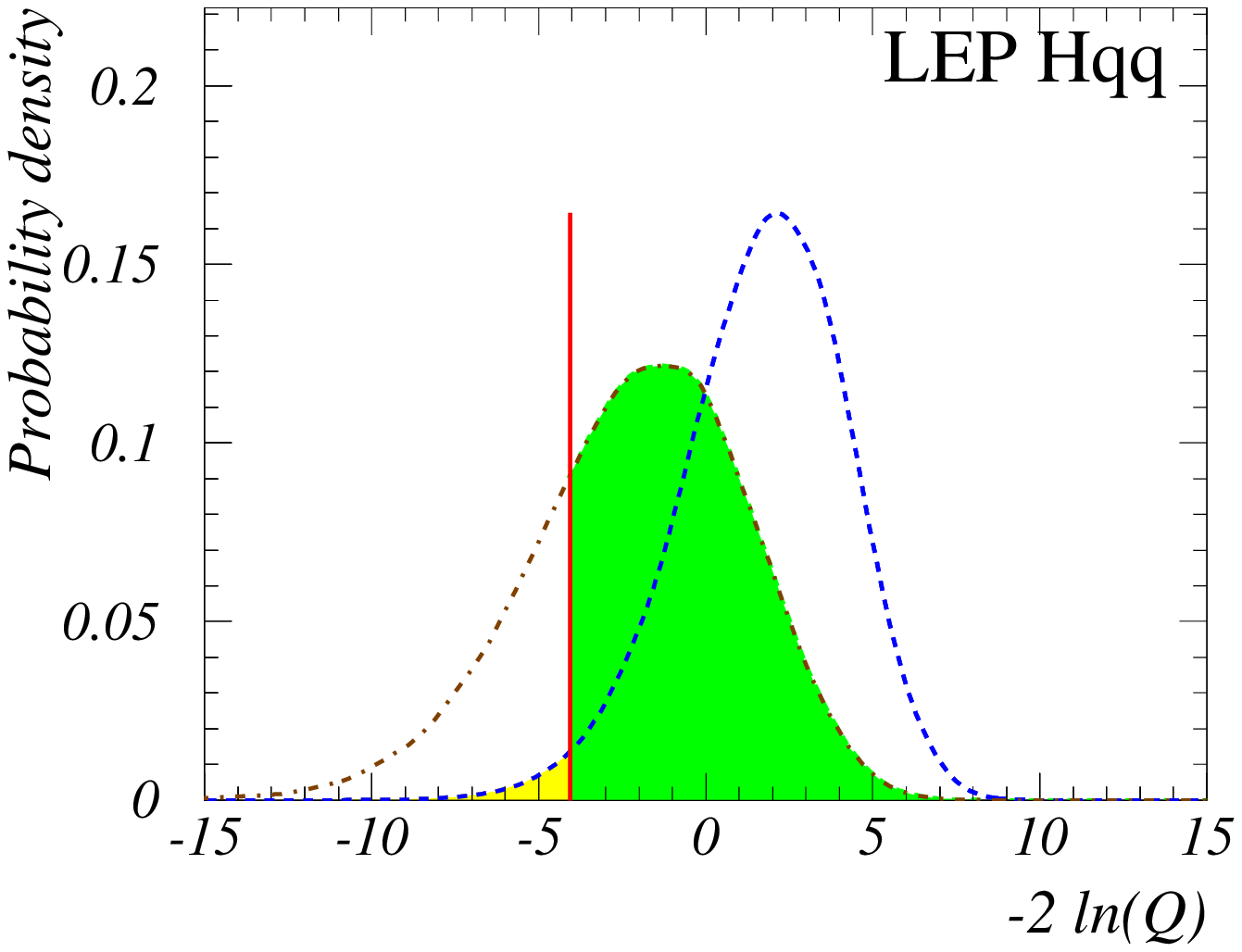,width=0.49\textwidth,height=4cm}
\epsfig{figure=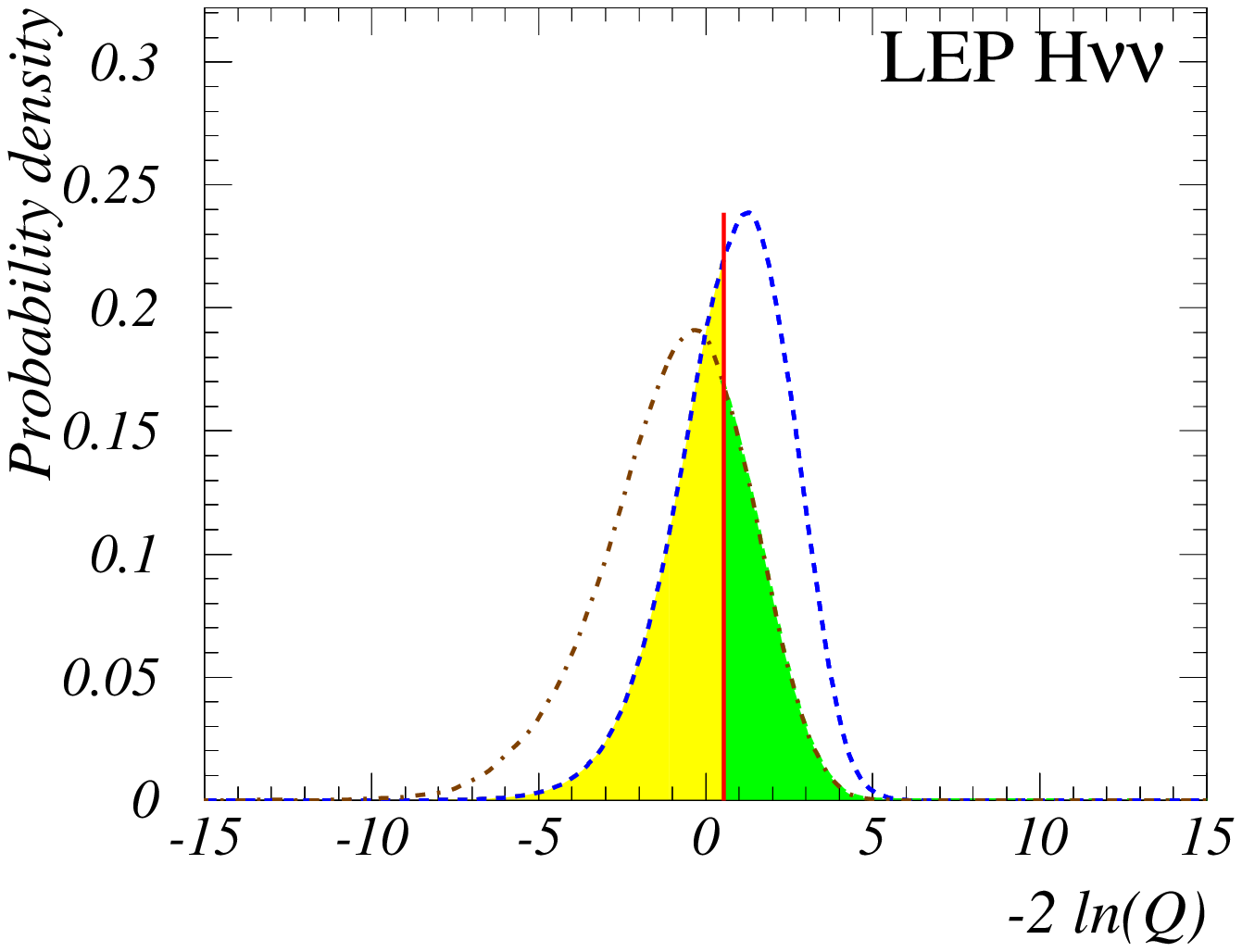,width=0.49\textwidth,height=4cm}\\
\epsfig{figure=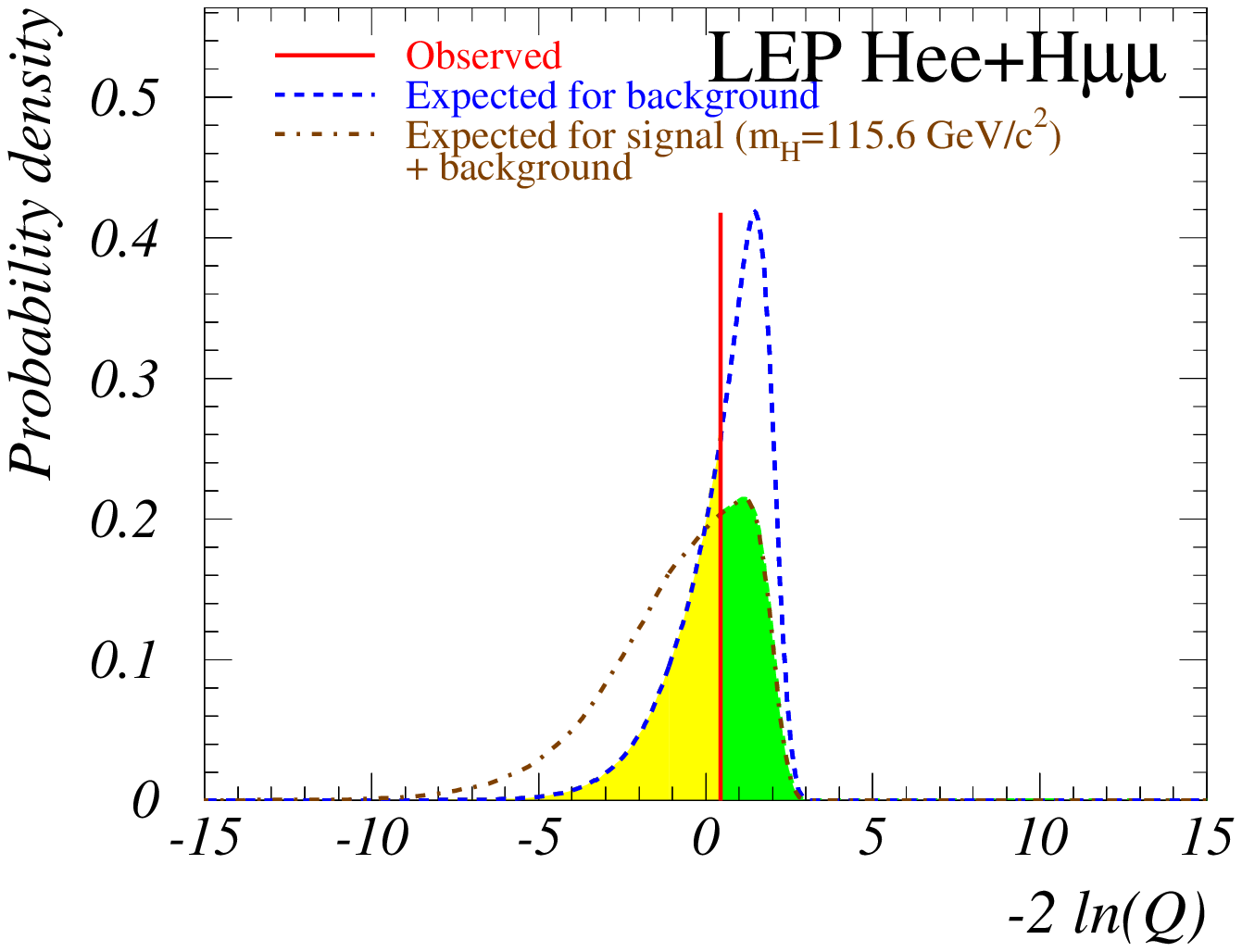,width=0.49\textwidth,height=4cm}
\epsfig{figure=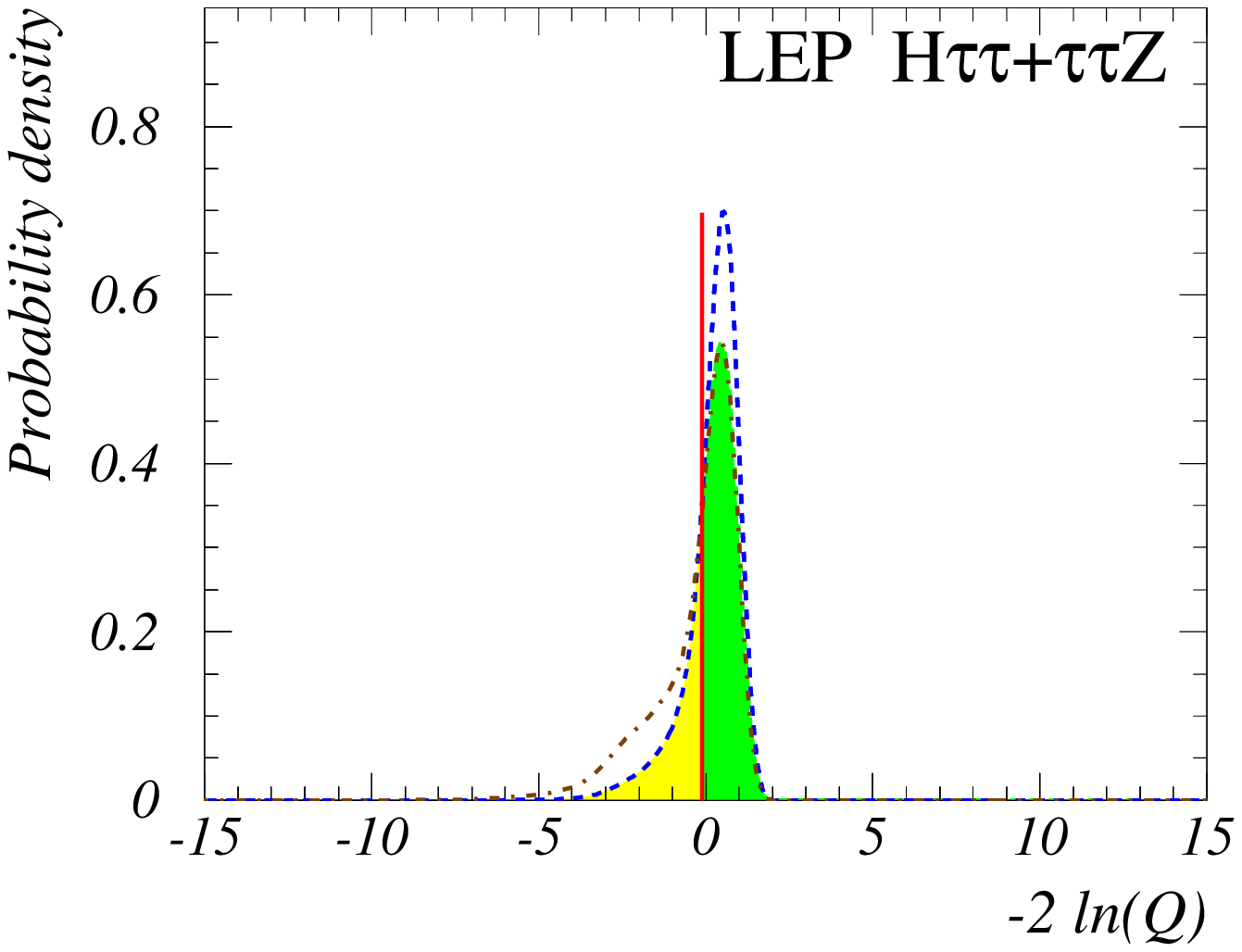,width=0.49\textwidth,height=4cm}\\
\epsfig{figure=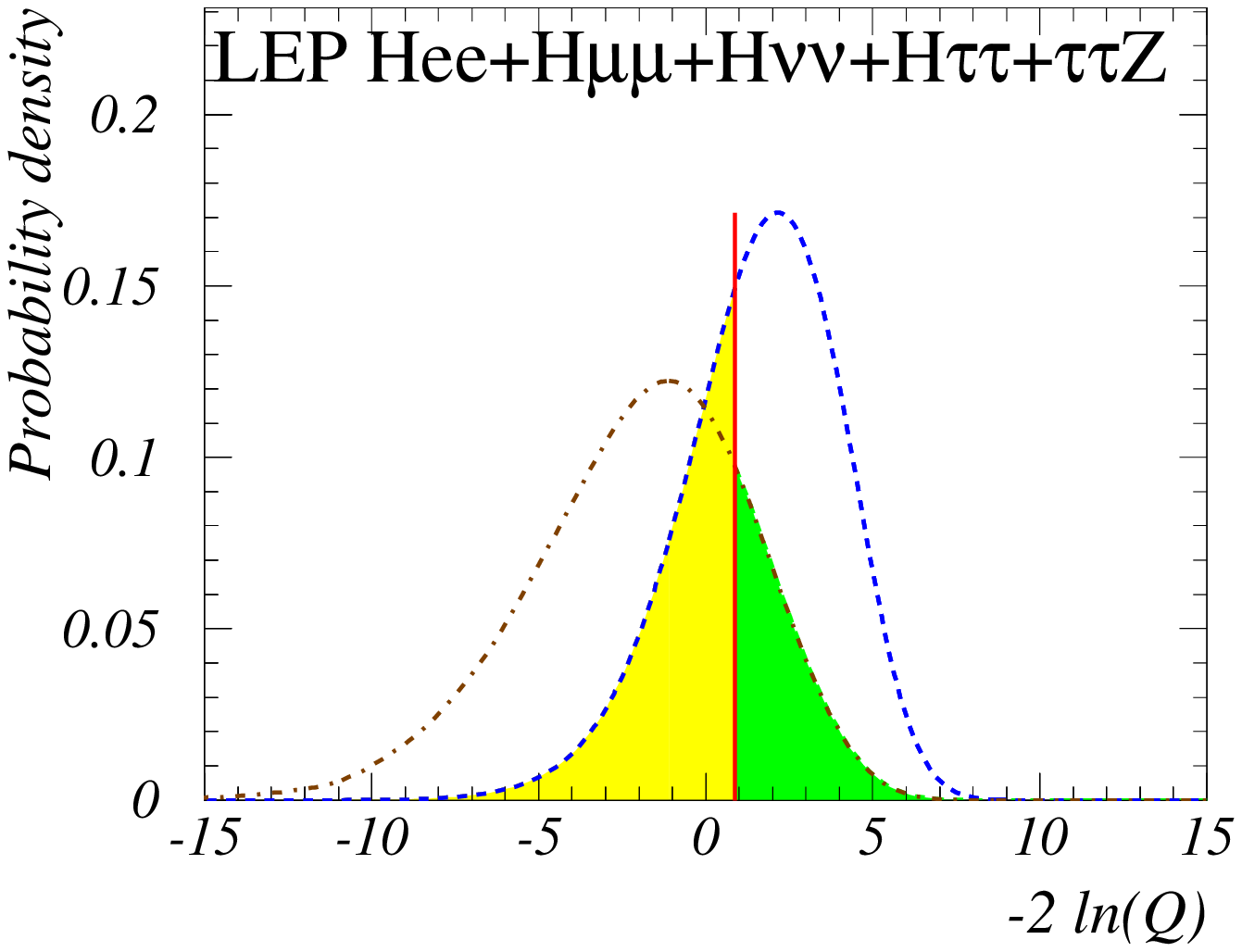,width=0.49\textwidth,height=4cm}\\
\caption[]{\small Probability density functions corresponding to a test-mass $m_H=115.6$~GeV 
for subsets of the data. Upper four plots: subdivision by experiments; 
next four plots: subdivision by decay channels.
The lowest plot combines all but the four-jet channel.
In each case, the observed value of $-2\ln Q$ is indicated by the vertical line.

\label{fig:indiv-prob-dens}}
\end{center}
\end{figure}

The first two subdivisions have been addressed 
in Figures~\ref{fig:a-d-l-o-lnq} and \ref{fig:channels-lnq} and Table~\ref{tab:clb}. The corresponding
probability density distributions for
$m_H=115.6$~GeV are shown in Figure~\ref{fig:indiv-prob-dens}.
The largest difference occurs between the subsets of ALEPH and DELPHI.
Looking separately at the final state topologies, the excess is mainly concentrating in the four-jet channel. 
Combining the four experiments while leaving out the four-jet channel, 
the lowest plot in Figure~\ref{fig:indiv-prob-dens} is obtained.
 
As seen in Figure~\ref{fig:adlo-weights}, the presence of a Higgs boson should affect
a substantial part of the event weight distribution. If the data set is subdivided in high- and low-purity subsets 
by selecting $s/b >1$ and $s/b <1$, at which point the two subsamples have approximately equal expected sensitivity, 
the contributions
to $-2\ln Q$ are consistent, and slightly more signal-like (negative) in the low-purity subset.
Hence, the observed excess is not due to a few events with exceptionally high weights only, but 
is reflected by the whole distribution of event weights.  

\section{Conclusion}
Combining the data from the four LEP experiments, 
a new lower bound for the mass of the Standard Model Higgs boson has been derived, which is 114.1~GeV at the 95\% confidence
level. There is an excess which can be interpreted as 
production of a Standard Model Higgs boson with a mass higher than the quoted limit. It is concentrated 
mainly in the data sets with centre-of-mass energies higher than 206~GeV. 
The likelihood test designates 115.6~GeV as the preferred mass. The probability for a 
fluctuation of the Standard Model background is 3.4\%. 
This effect is mainly driven by the ALEPH data and the four-jet final state.

\begin{center}
{THE RESULTS QUOTED IN THIS PAPER ARE NOT FINAL \\
SINCE THEY COMBINE PRELIMINARY RESULTS FROM THREE EXPERIMENTS\\
WITH FINAL RESULTS FROM ONE EXPERIMENT}
\end{center}
\vspace{2cm}
ACKNOWLEDGEMENTS\\

\noindent 
We congratulate our colleagues from the LEP Accelerator Division for the successful running in the year 2000 at 
the highest energies, and would like to express our thanks
to the engineers and technicians in all our institutions for their contributions to the excellent
performance of the four LEP experiments. The LEP Higgs working group acknowledges the fruitful cooperation between the
experiments in developing the combination procedures and in putting them into application.


\newpage

\end{document}